\documentclass[notitlepage,nofootinbib,aps,12pt,tightenlines,%
  showpacs,showkeys]{revtex4}

\usepackage{amsmath}
\usepackage{amssymb,amsfonts}
\usepackage{bm}
\usepackage[mathcal]{euscript}
\usepackage{graphicx}
\usepackage{psfrag}
\usepackage{subfigure}

\numberwithin{equation}{section}

\graphicspath{{figs/}{figs_lo/}}

\newcommand{\comment}[1]{}
\newcommand{\savecomment}[1]{}

\newcommand{\ie}{\textit{i.e.}}
\newcommand{\etal}{\textit{et~al.}}
\newcommand{\eg}{\textit{e.g.}}

\newcommand{\mathnotation}[2]{\newcommand{#1}{\ensuremath{#2}}}

\newcommand{\goodgap}{%
	\hspace{\subfigtopskip}%
	\hspace{\subfigbottomskip}}

\newcommand{\pow}[2]{\{#1\}^#2}

\renewcommand{\l}{\left}			
\renewcommand{\r}{\right}			
\mathnotation{\pd}{\partial}			
\mathnotation{\ldef}{\mathrel{\raisebox{.069ex}{:}\!\!=}}
\mathnotation{\rdef}{\mathrel{=\!\!\raisebox{.069ex}{:}}}
\mathnotation{\dint}{\,{\mathrm{d}}}		
\newcommand{\Order}[1]{\ensuremath{\mathcal{O}\!\l(#1\r)}}

\mathnotation{\grad}{\nabla}			
\renewcommand{\div}{\grad\cdot}			
\mathnotation{\curl}{\grad\times}		
\mathnotation{\lapl}{\Delta}			

\renewcommand{\time}{\tau}			
\mathnotation{\xc}{x}				
\mathnotation{\xv}{\bm{x}}			
\mathnotation{\yc}{y}				
\mathnotation{\uc}{u}				
\mathnotation{\uct}{\tilde{u}}			
\mathnotation{\ucb}{\bar{u}} 			
\mathnotation{\uv}{{\bm{u}}}			
\mathnotation{\vc}{v}				
\mathnotation{\vct}{\tilde{v}}			
\mathnotation{\vcb}{\bar{v}}			
\newcommand{\ev}[1]{\bm{e}_#1}			
\newcommand{\evup}[1]{\bm{e}^#1}		
\newcommand{\euv}[1]{\hat{\bm{e}}_#1}		
\newcommand{\evt}[1]{\tilde{\bm{e}}_#1}		
\newcommand{\evtup}[1]{\tilde{\bm{e}}^#1}	
\newcommand{\evb}[1]{\bar{\bm{e}}_#1}		

\mathnotation{\grav}{g}				
\mathnotation{\gravy}{\hat\grav_\yc}		
\mathnotation{\gravsc}{{\hat\grav_{\mathrm{s}}}}
\mathnotation{\gravsv}{\hat{\bm{\grav}}_{\mathrm{s}}}
\mathnotation{\gravv}{\hat{\bm{\grav}}}		
\mathnotation{\pres}{p}				
\mathnotation{\prest}{\tilde{p}}		
\mathnotation{\visc}{\nu}			
\mathnotation{\stress}{\mathbb{T}}		
\mathnotation{\Qc}{q}				
\mathnotation{\Qct}{\tilde{\Qc}}		
\mathnotation{\Qcb}{\bar{\Qc}}			
\mathnotation{\Qv}{\bm{\Qc}}			

\mathnotation{\Metricc}{\mathbb{G}}		
\mathnotation{\metricc}{\mathfrak{g}}		
\mathnotation{\Xc}{X}				
\mathnotation{\Xv}{\bm{\Xc}}			
\mathnotation{\rc}{r}				
\mathnotation{\rv}{\bm{\rc}}			
\mathnotation{\arf}{\omega}			
\mathnotation{\arft}{\tilde{\arf}}		
\mathnotation{\arfb}{\bar{\arf}}		

\mathnotation{\kcurv}{k}			
\mathnotation{\Gcurv}{\mathcal{G}}		
\mathnotation{\kappasurf}{\kappa_{\mathrm{surf}}}

\mathnotation{\AAA}{A}
\mathnotation{\BBB}{B}
\mathnotation{\aaa}{a}
\mathnotation{\bbb}{b}
\mathnotation{\ccc}{c}
\mathnotation{\ddd}{d}

\mathnotation{\eps}{\varepsilon}

\mathnotation{\fmg}{f}				
\mathnotation{\pmg}{p}				
\mathnotation{\qmg}{q}				
\mathnotation{\rmg}{r}				
\mathnotation{\smg}{s}				
\mathnotation{\tmg}{t}				
\mathnotation{\wmg}{w}				
\mathnotation{\wmgt}{\tilde{\wmg}}		%
\mathnotation{\wmgb}{\bar{\wmg}}		%

\begin{document}

\title{Transport in Thin Gravity-driven Flow \\
  over a Curved Substrate}
\author{Jean-Luc Thiffeault}
\email{jeanluc@imperial.ac.uk}
\author{Khalid Kamhawi}
\affiliation{Department of Mathematics, Imperial College London,
SW7 2AZ, United Kingdom}
\date{31 July 2006}

\begin{abstract}
We consider steady gravity-driven flow of a thin layer of viscous fluid over a
curved substrate.  The substrate has topographical variations (`bumps') on a
large scale compared to the layer thickness.  Using lubrication theory, we
find the velocity field in generalized curvilinear coordinates. We correct the
velocity field so as to satisfy kinematic constraints, which is essential to
avoid particles escaping the fluid when computing their trajectories.  We then
investigate the particle transport properties of flows over substrates with
translational symmetry, where chaotic motion is precluded.  The existence of
trapped and untrapped trajectories leads to complicated transport properties
even for this simple case.  For more general substrate shapes, the
trajectories chaotically jump between trapped and untrapped motions.
\end{abstract}


\maketitle

\section{Introduction}

It is often the case in applications that viscous fluid flows in a thin layer
at low Reynolds number, which allows lubrication (or long-scale, also
long-wave) theory to be used.~\cite{Oron1997} The resulting decoupling between
the long-scale and the `thin' direction greatly simplifies the mathematical
solution of the flow.  There is essentially only one equation to be solved,
for the thickness of the fluid as a function of position.  This equation
arises from mass conservation appropriately averaged over the thin coordinate.
The velocity is then deduced from the thickness.

The mass-conservation equation to be solved involves only the long-scale
coordinates.  The third coordinate is `slaved' to the horizontal ones by the
effect of viscosity.  In that sense one usually thinks of thin-layer flows as
`two-dimensional,' in the sense that vertical variations (\ie, in the thin
direction) exist but are subsumed in the theory and that the vertical velocity
field is small.  However, if one is interested in the transport properties of
such flows, any amount of vertical variation is crucial.  We will show here
that even small variations can lead to chaotic behavior of particle
trajectories, known as chaotic advection,~\cite{Aref1984,Ottino,Wiggins2004}
greatly enhancing the transport properties of the flow.  Indeed, if the flow
were truly two-dimensional and steady, chaotic transport would be impossible
since two-dimensional autonomous dynamical systems cannot exhibit
chaos.~\cite{Eckmann1985}

As far as we know, the transport properties of thin-layer flows have not been
studied, beyond plotting streamlines in two-dimensional cases, probably
because they were assumed too simple to be important.  But understanding these
properties could be important if, for instance, a compositionally-uniform
coating is required in an industrial application.  A related area that has
received much attention lately is chaotic advection in
microchannels.~\cite{Stroock2002} There periodic grooved patterns at the
bottom of the channel cause swirl in the flow that leads to chaos.  Here we
study a different regime where the bottom substrate has larger topographical
variations, the is fluid thinner, and the sidewalls less important.

We will show here that the shape of the curved substrate is crucial.  Though
we investigate this for the case of steady gravity-driven flow with a free
surface, the conclusions reached should qualitatively apply to different
thin-flow situations such as those driven by surface tension gradients.
Constant surface tension is included in the derivation, but left out of most
of the examples since it does not affect the qualitative results.

There are several derivations of models of thin fluid flow over curved
substrates in the literature.~\cite{Ida1998,Kalliadasis2000,%
Kalliadasis2001,Mazouchi2001,Roy2002,Myers2002} Our approach follows that of
Roy~\etal~\cite{Roy2002} (hereafter RRS), in that the vertical
(thin-direction) coordinate is orthogonal to the long-scale coordinates, but
differs in that the long-scale coordinates are not necessarily orthogonal to
each other.  This is because even though orthogonal coordinates can always be
found,~\cite{Kreyszig} doing so for a general substrate (other than planes,
cylinders, and spheres) requires solving differential equations numerically.
Working in generalized coordinates is not much more complicated if
differential-geometric notation is used,~\cite{Wald,Synge,Schutz} and allows a
simple parametrization of the substrate in terms of a height function.

Another motivation for re-deriving the model of RRS~\cite{Roy2002} is that we
need to include correction terms to the mass-conservation equation and
vertical velocity to ensure that the kinematic boundary condition at the free
surface is satisfied exactly, up to numerical accuracy.  In the RRS model, the
kinematic boundary condition is satisfied to second order in the thickness.
This is fine for most applications, but for particle advection it is
catastrophic: in numerical simulations, individual particles will gleefully
escape the top surface if they come anywhere near it, since even a
second-order error is enough to give a small component of the velocity field
normal to the surface.  Rather than remedy this in some ad-hoc manner (by
reflecting the particles back in, for instance), we instead include some
judiciously-chosen second-order terms that guarantee that the kinematic
boundary condition at the top surface is satisfied.

The outline of the paper is as follows.
Sections~\ref{sec:coords}--\ref{sec:dynamics} comprise the derivation of the
model equations.  At each stage, we carefully record where the thinness of the
layer comes in as an approximation.  Since we will be working with substrates
of complicated shapes, the first part of of our paper
(Section~\ref{sec:coords}) is devoted to establishing a convenient coordinate
system for our problem.  Then we tackle mass conservation in those coordinates
in Section~\ref{sec:masscons}.  Section~\ref{sec:dynamics} introduces the
dynamical equations of motion for Stokes flow.  In that section we also
rescale the variables (Section~\ref{sec:small}) and solve for the velocity
field (Section~\ref{sec:vel}) to first order in the thickness.  This gives the
mass flux vector, whose divergence must vanish for steady flow.  We include
second-order terms to the mass flux to impose the kinematic boundary condition
at the free surface, which is essential for particle advection as discussed
above.

In Section~\ref{sec:PDEsolve} we discuss solutions of the governing PDE for
the fluid thickness.  The simplest type of solution involves substrates with a
translational symmetry, for which the steady flow can be found analytically at
leading order.  However, solving for the flow at first order, or for
substrates without such a symmetry, we must resort to numerical solutions.

Section~\ref{sec:advection} is concerned with fluid particle advection and is
the central results section of the paper.  We examine the variety of particle
trajectories by making Poincar\'e surfaces of section, for increasingly
complicated substrates.  As expected, the transport properties also become
more complicated, exhibiting a range of behavior from trapped to chaotic
trajectories.  We offer some concluding remarks in
Section~\ref{sec:discussion}.

\section{Coordinate System}
\label{sec:coords}

\subsection{Separating the Thin Direction}
\label{sec:thindir}

In our problem, fluid motion occurs over a curved substrate of arbitrary
shape.  The direction normal to the substrate is special in that it defines
the direction in which the fluid layer is assumed `thin.'  Hence, it is
convenient to locate a point~$\rv$ in the fluid as
\begin{equation}
  \rv(\xc^1,\xc^2,\yc) = \Xv(\xc^1,\xc^2) + \yc\,\euv{3}(\xc^1,\xc^2)
  \label{eq:r}
\end{equation}
where~$\Xv(\xc^1,\xc^2)$ is the location of the substrate,~$\euv{3}$ is a unit
vector normal to the substrate, and~$\yc$ is the perpendicular distance
from~$\rv$ to the substrate.  The coordinates~$\xc^1$ and~$\xc^2$ are
substrate coordinates used to localize points on the substrate.  For example,
in Section~\ref{sec:monge} we will use the Monge
parametrization,~$\Xv=\begin{pmatrix}\xc^1 & \xc^2 &
\fmg(\xc^1,\xc^2)\end{pmatrix}^T$, where~$\fmg$ gives the height of the
substrate.

The tangent vectors to the substrate are
\begin{equation}
  \ev{\alpha} = \pd_\alpha\Xv
  \label{eq:tangsubstr}
\end{equation}
where~$\pd_\alpha \ldef \pd/\pd\xc^\alpha$.  The coordinate vectors
associated with the coordinate system are found from~\eqref{eq:r},
\begin{equation}
  \evt{\alpha} = \pd_\alpha\rv = \ev{\alpha}
  - \yc\, {\mathbb{K}_\alpha}^\beta\,\ev{\beta}\,,\qquad
  \frac{\pd\rv}{\pd \yc} = \euv{3}\,,
  \label{eq:tang}
\end{equation}
where we assume the summation of repeated indices, and Greek indices only
take the value~$1$ or~$2$.  We have also defined the curvature
tensor~${\mathbb{K}_\alpha}^\beta$ from
\begin{equation}
  \pd_\alpha\euv{3} = -{\mathbb{K}_\alpha}^\beta\,\ev{\beta}\,,
  \label{eq:curv}
\end{equation}
since changes in the unit vector~$\euv{3}$ must be perpendicular to~$\euv{3}$,
and~$\euv{3}\cdot\ev{\alpha}=0$.  Note that the~$\ev{\alpha}$ are not
necessarily orthogonal or normalised.

We adopt the convention that quantities with a tilde are evaluated in the
`bulk' (away from the substrate), and thus depend on~$\yc$, whilst those
without the tilde are `substrate' quantities and do not depend on~$\yc$.
Thus,~$\evt{\alpha}(\xc^1,\xc^2,0)=\ev{\alpha}(\xc^1,\xc^2)$.  Starting in
Section~\ref{sec:masscons}, we will also denote quantities evaluated on the
free surface at~$\yc=\eta$ by an overbar.

The three-dimensional metric tensor~$\tilde\metricc_{ab}$ has components
\begin{equation*}
  \tilde\metricc_{\alpha\beta} = \evt{\alpha}\cdot\evt{\beta} =
  \widetilde\Metricc_{\alpha\beta}\,,\quad
  \tilde\metricc_{\alpha3} = \evt{\alpha}\cdot\euv{3} = 0\,,\quad
  \tilde\metricc_{33} = \euv{3}\cdot\euv{3} = 1\,,
\end{equation*}
where
\begin{equation}
\begin{split}
  \widetilde\Metricc_{\alpha\beta} &\ldef \evt{\alpha}\cdot\evt{\beta} =
  \Metricc_{\alpha\beta} - 2\yc\,\mathbb{K}_{\alpha\beta}
  + \pow{\yc}{2}\,{\mathbb{K}_\alpha}^\gamma\,\mathbb{K}_{\gamma\beta}\,,\\
  \Metricc_{\alpha\beta} &\ldef \ev{\alpha}\cdot\ev{\beta}\,,
\end{split}
\label{eq:GtG}
\end{equation}
and we have written the symmetric tensor~$\mathbb{K}_{\alpha\beta} =
{\mathbb{K}_\alpha}^\gamma\,\Metricc_{\beta\gamma}$ in the usual manner of
lowering indices with the metric tensor.~\cite{Wald,Synge,Schutz} The
three-dimensional metric tensor is thus block-diagonal, and the~$\yc$
coordinate is unstretched compared to the Cartesian coordinate system.  It
measures the true perpendicular distance from the substrate to a point in the
fluid.  Note that it is important that indices be raised or lowered with the
appropriate metric: $\Metricc_{\alpha\beta}$ for substrate quantities,
$\tilde\metricc_{ab}$ or $\widetilde\Metricc_{\alpha\beta}$ for bulk
quantities.  The tensor $\widetilde\Metricc_{\alpha\beta}(\xc^1,\xc^2,\yc)$
can be used to raise or lower indices, even though it depends on three
coordinates but only has two-dimensional indices, because of the
block-diagonality of the metric~$\tilde\metricc_{ab}$.  Note also that in
Eq.~\eqref{eq:GtG} we have written~$\pow{\yc}{2}$ rather than~$\yc^2$, since
the second form could be misinterpreted as the second contravariant component
of~$\yc$ (which is nonsense in this case).  We shall follow this practice
throughout the paper, but note that negative powers are written as~$\yc^{-2}$,
since then there is no confusion.

Given the substrate vectors~$\ev{\alpha}$, it is easy to solve for the
covectors~$\evup{\alpha}$, which are such
that \hbox{$\evup{\alpha}\cdot\ev{\beta}={\delta_\beta}^\alpha$}.  Then the
bulk covectors are
\begin{equation}
  \evtup{\alpha} = \evup{\alpha}
  + \yc\,{\mathbb{K}_\beta}^\alpha\,\evup{\beta} + \Order{\pow{\yc}{2}},
  \label{eq:cotang}
\end{equation}
to first order in~$\yc$, which can be checked by taking their dot product with
the coordinate vectors~\eqref{eq:tang} to see that
\hbox{$\evtup{\alpha}\cdot\evt{\beta}={\delta_\beta}^\alpha +
\Order{\pow{\yc}{2}}$}.  We will not need the covectors to higher order
in~$\yc$.  From~\eqref{eq:cotang}, we find the inverse metrics,
\begin{equation}
\begin{split}
  \widetilde\Metricc^{\alpha\beta} &= \evtup{\alpha}\cdot\evtup{\beta}
  = \Metricc^{\alpha\beta}
  + 2\yc\,\mathbb{K}^{\alpha\beta} +
  \Order{\pow{\yc}{2}},\label{eq:Metrictinv}\\
  \Metricc^{\alpha\beta} &= \evup{\alpha}\cdot\evup{\beta}\,,
\end{split}
\end{equation}
where again the higher-order terms will not be needed.

A crucial quantity is the determinant of the metric,
\begin{equation}
  \wmgt = \bigl(\det\,\widetilde\Metricc_{\alpha\beta}\bigr)^{1/2}
  = \arft\,\wmg,\qquad
  \wmg = \bigl(\det\,\Metricc_{\alpha\beta}\bigr)^{1/2}\,,
  \label{eq:wdef}
\end{equation}
where
\begin{equation}
  \arft = 1 - \kappa\,\yc + \Gcurv\,\pow{\yc}{2}\,,
  \label{eq:arf}
\end{equation}
and
\begin{subequations}
\begin{alignat}{2}
  \kappa &= {\mathbb{K}_\alpha}^\alpha\,\qquad
  &&\text{mean curvature}\label{eq:meancurv}\\
  \Gcurv &= \det\,{\mathbb{K}_\alpha}^\beta\,\qquad
  &&\text{Gaussian curvature}\,.\label{eq:Gausscurv}
\end{alignat}
\end{subequations}
The area element at fixed~$\yc$ is given by~$\arft\,\wmg\,d\xc^1d\xc^2$, and
the volume element by~$\arft\,\wmg\,d\xc^1d\xc^2d\yc$, since the~$\yc$
coordinate is unstretched with respect to the Cartesian system.  The volume
per unit substrate area of a fluid layer of thickness~$\eta$ is thus
\begin{equation}
  \zeta =
  \frac{\int_0^\eta\arft\,\wmg\,d\xc^1d\xc^2d\yc}{\wmg\,d\xc^1d\xc^2}
  = \eta - \tfrac{1}{2}\kappa\,\pow{\eta}{2} +
  \tfrac{1}{3}\Gcurv\,\pow{\eta}{3}\,.
  \label{eq:zeta}
\end{equation}
The quantity~$\arft$ also tells us when the coordinate system ceases to be
valid: if~$\arft=0$, then the area element becomes singular.  Hence, the valid
range for~$\yc$ is
\begin{equation*}
  0 \le \yc < \{\max(\kcurv_1,\kcurv_2,0^+)\}^{-1}\,,\qquad
  \kcurv_{1,2} \ldef
  \tfrac{1}{2}\bigl(\kappa \pm \sqrt{\pow{\kappa}{2} - 4\Gcurv}\bigr)
\end{equation*}
where~$\kcurv_1$ and~$\kcurv_2$ are the principal curvatures.~\cite{Kreyszig}
If both principal curvatures are negative then the positive range in~$\yc$ is
unrestricted (convex region); otherwise~$\yc$ is limited by the largest
positive principal curvature.

We now have all the geometrical equipment we need to solve fluid equations on
the curved substrate.  Note that all the results of this section are exact,
except for the covectors~\eqref{eq:cotang} and inverse
metric~\eqref{eq:Metrictinv}, which are valid to first order in~$\yc$.

\subsection{Substrate Coordinates}
\label{sec:monge}

For most applications in the literature of thin films, orthonormal coordinates
have been the coordinates of choice.  This is because the main substrate
shapes that have been treated are planes, cylinders, and spheres, where
orthonormal coordinates are readily available.  For a general substrate shape,
orthonormal coordinates are difficult to construct and require numerical
integration.  Singularities (umbilics) also cause problems.~\cite{Kreyszig}
For our application---flow down a curved substrate---the Monge representation
of a surface~\cite{Flanders} is the most convenient.

The Monge representation is a glorified name for a parametrization of the
substrate by
\begin{equation}
  \Xv(\xc^1,\xc^2)
  =\begin{pmatrix}\xc^1 & \xc^2 & \fmg(\xc^1,\xc^2)\end{pmatrix}^T
  \label{eq:Xsurf}
\end{equation}
in three-dimensional Cartesian space.  Following standard
notation,~\cite{Flanders} we define
\begin{subequations}
\begin{gather}
  \pmg = \pd_1\fmg,\quad
  \qmg = \pd_2\fmg,\quad\\
  \rmg = \pd_1\pd_1\fmg,\quad
  \smg = \pd_1\pd_2\fmg,\quad
  \tmg = \pd_2\pd_2\fmg.
\end{gather}%
\label{eq:mongeparams}%
\end{subequations}%
The unnormalized, nonorthogonal tangents~$\ev{1}$ and~$\ev{2}$ are
\begin{equation*}
  \ev{1} = \pd_1\Xv
  = \begin{pmatrix}1 & 0 & \pmg\end{pmatrix}^T,\qquad
  \ev{2} = \pd_2\Xv
  = \begin{pmatrix}0 & 1 & \qmg\end{pmatrix}^T,
\end{equation*}
and their normalised cross product gives the normal to the substrate,
\begin{equation*}
  \euv{3} = \frac{1}{\wmg}\,\begin{pmatrix}-\pmg & -\qmg & 1\end{pmatrix}^T.
\end{equation*}
The corresponding covectors are
\begin{equation*}
  \evup{1} = \frac{1}{\pow{\wmg}{2}}\begin{pmatrix}
    (1+\pow{\qmg}{2}) & -\pmg\qmg & \pmg\end{pmatrix}^T,\qquad
  \evup{2} = \frac{1}{\pow{\wmg}{2}}\begin{pmatrix}-\pmg\qmg
    & (1+\pow{\pmg}{2}) & \qmg\end{pmatrix}^T
\end{equation*}
and~$\euv{3}$ is its own covector.  The metric tensor of the substrate and its
inverse are
\begin{equation}
  \{\Metricc_{\alpha\beta}\} = \ev{\alpha}\cdot\ev{\beta}
  = \begin{pmatrix}1+\pow{\pmg}{2} & \pmg\qmg \\ \pmg\qmg &
  1+\pow{\qmg}{2}\end{pmatrix},
    \quad
  \{\Metricc^{\alpha\beta}\} = \evup{\alpha}\cdot\evup{\beta}
  = \frac{1}{\pow{\wmg}{2}}
  \begin{pmatrix}1+\pow{\qmg}{2} & -\pmg\qmg \\ -\pmg\qmg &
  1+\pow{\pmg}{2}\end{pmatrix},
    \label{eq:GMonge}
\end{equation}
with determinant
\begin{equation*}
  \wmg = \bigl(\det\,\Metricc_{\alpha\beta}\bigr)^{1/2}
  = \sqrt{1 + \pow{\pmg}{2} + \pow{\qmg}{2}}\,.
\end{equation*}
The curvature tensor is
\begin{equation}
  \{{\mathbb{K}_\alpha}^\beta\} \ldef \frac{1}{\pow{\wmg}{3}}\begin{pmatrix}
    (1+\pow{\qmg}{2})\rmg - \pmg\qmg\smg & (1+\pow{\pmg}{2})\smg -
    \pmg\qmg\rmg \\
    (1+\pow{\qmg}{2})\smg - \pmg\qmg\tmg & (1+\pow{\pmg}{2})\tmg - \pmg\qmg\smg
    \label{eq:Kdefmg}
  \end{pmatrix}
\end{equation}
which gives mean and Gaussian curvatures
\begin{equation*}
\begin{split}
  \kappa &= {\mathbb{K}_\alpha}^\alpha
  =\frac{1}{\pow{\wmg}{3}}\l((1+\pow{\qmg}{2})\rmg - 2\pmg\qmg\smg
  + (1+\pow{\pmg}{2})\tmg\r),\\
  \Gcurv &= \det\,{\mathbb{K}_\alpha}^\beta
  = \frac{1}{\pow{\wmg}{4}}\l(\rmg\tmg - \pow{\smg}{2}\r).
\end{split}
\end{equation*}

We write the normalized gravity vector as~$\gravv = (\sin\theta\cos\phi\ \
\sin\theta\sin\phi\ \ -\cos\theta)^T$, so that the inclination angle~$\theta$
is zero for gravity pointing downwards, and for~$\phi\in(-\pi/2,\pi/2)$
positive~$\theta$ induces flow in the positive~$\xc^1$ direction.  Then we
have the components
\begin{equation}
\begin{split}
\gravsc^1 &= \gravv\cdot\evup{1}
  = -\l(\pmg\,\cos\theta + \pmg\qmg\,\sin\theta\sin\phi
  - (1+\pow{\qmg}{2})\sin\theta\cos\phi\r)/\pow{\wmg}{2}\,,\\
\gravsc^2 &= \gravv\cdot\evup{2}
  = -\l(\qmg\,\cos\theta + \pmg\qmg\,\sin\theta\cos\phi
  - (1+\pow{\pmg}{2})\sin\theta\sin\phi\r)/\pow{\wmg}{2}\,,\\
\gravy &= \gravv\cdot\euv{3}
  = -\l(\cos\theta + \pmg\,\sin\theta\cos\phi
  + \qmg\,\sin\theta\sin\phi\r)/\wmg\,.
\end{split}%
\label{eq:gravvec}%
\end{equation}%
The specific parametrization of the substrate introduced in this section will
not be needed in the derivation of the equations of motion
(Sections~\ref{sec:masscons} and~\ref{sec:dynamics}), only in their solution
(Section~\ref{sec:gensol}).  Hence, a different parametrization could be used
if called for by the geometry of the substrate.  For instance, flow down a
curved filament is better parametrized by cylindrical coordinates, or if the
substrate has overhangs (making~$\fmg$ multivalued) coordinates based on arc
length are better.

\section{Mass Conservation}
\label{sec:masscons}

We remind the reader of our notational convention:
\begin{enumerate}
\item Quantities with a \textbf{tilde} (\eg, $\evt{\alpha}$ and~$\wmgt$) are
  evaluated between the substrate and the free surface (in the `bulk') and are
  functions of~$(\xc^1,\xc^2,\yc,\time)$.
\item Quantities with an \textbf{overbar} (\eg, $\evb{\alpha}$ and~$\wmgb$)
  are evaluated on the free surface~\hbox{$\yc=\eta(\xc^1,\xc^2,\time)$} and
  are functions of~$(\xc^1,\xc^2,\time)$.
\item `Bare-headed' quantities (\eg, $\ev{\alpha}$ and~$\wmg$) are evaluated
  on the substrate~$\yc=0$ and are functions of~$(\xc^1,\xc^2)$, or they are
  quantities that do not depend on~$\yc$ at all (\eg, $\euv{3}$).
\end{enumerate}

We introduce a velocity field
\begin{equation}
  \uv = \uct^\alpha\,\evt{\alpha} + \vct\,\euv{3}\,,
  \label{eq:vel}
\end{equation}
associated with some time-dependent incompressible flow.  Here we depart from
RRS~\cite{Roy2002} in two ways: first, the $\evt{\alpha}$ are not unit
vectors, so the velocity components~$\uct^\alpha$ are scaled differently;
second, we use $\evt{\alpha}$ and not~$\ev{\alpha}$, so that the velocity
field is expressed in terms of the bulk coordinate vectors rather than the
substrate tangent vectors.  This is slightly more natural in the present
context, but is not crucial.

Mass conservation is imposed via the divergence-free
condition,~$\div\uv = 0$; in terms of the coordinates in
section~\ref{sec:coords}, this is
\begin{equation}
  \pd_\alpha\l(\wmgt\,\uct^\alpha\r)
  + \frac{\pd}{\pd\yc}\l(\wmgt\,\vct\r) = 0\,.
  \label{eq:divucov}
\end{equation}
We integrate this from~$0$ to~$\eta$ and use the no-throughflow
condition~$\vct(\xc^1,\xc^2,0)=0$ to get
\begin{align}
  \wmgb\,\vcb &= -\int_0^\eta\pd_\alpha\l(\wmgt\,\uct^\alpha\r)\dint\yc\\
  &= \wmgb\,\ucb^\alpha\pd_\alpha\eta
  -\pd_\alpha\int_0^\eta\l(\wmgt\,\uct^\alpha\r)\dint\yc.
\end{align}
Now we use the kinematic boundary condition at the top surface,
\begin{equation}
  \pd_\time\eta + \ucb^\alpha\,\pd_\alpha\eta = \vcb,
  \label{eq:kinematicBC}
\end{equation}
to find
\begin{equation}
  \wmgb\,\pd_\time\eta =
  -\pd_\alpha\int_0^\eta\l(\wmgt\,\uct^\alpha\r)\dint\yc.
  \label{eq:interm1}
\end{equation}
We define the mass flux vector \savecomment{Current?}
\begin{equation}
  \Qct^\alpha(\xc^1,\xc^2,\yc)
  \ldef \int_0^\yc\arft\,\uct^\alpha d\yc\,,\qquad
  \Qcb^\alpha(\xc^1,\xc^2) = \Qct^\alpha(\xc^1,\xc^2,\eta)
  \label{eq:flux}
\end{equation}
which allows us to rewrite~\eqref{eq:interm1} as
\begin{equation*}
  \wmgb\,\pd_\time\eta = -\pd_\alpha\l(\wmg\,\Qcb^\alpha\r).
\end{equation*}
We divide through by~$\wmg$,
\begin{equation}
  \arfb\,\pd_\time\eta = -\grad_\alpha\Qcb^\alpha
  \label{eq:almosthere}
\end{equation}
where the covariant divergence~$\grad_\alpha\Qcb^\alpha \ldef
\wmg^{-1}\pd_\alpha(\wmg\,\Qcb^\alpha)$, with~$\grad_\alpha$ denoting the
covariant derivative with respect to~$\xc^\alpha$ using the substrate
metric~$\Metricc_{\alpha\beta}$.~\cite{Wald,Synge,Schutz} In
Eq.~\eqref{eq:almosthere},~$\arft$ defined by Eq.~\eqref{eq:arf} is evaluated
at~$\yc=\eta$, and~$\time$ only enters~$\arft$ through~$\eta$.  Hence, we can
rewrite~\eqref{eq:almosthere} as
\begin{equation}
  \frac{\pd\zeta}{\pd\time}
  + \grad_\alpha\Qcb^\alpha = 0,
  \label{eq:masscons}
\end{equation}
where~$\zeta$ is the volume per unit substrate area given by
Eq.~\eqref{eq:zeta}.  Equation~\eqref{eq:masscons} thus embodies mass
conservation.

Note that there were no assumptions on the thinness of the layer in this
section.

\section{Dynamical Equations of Motion}
\label{sec:dynamics}

Sections~\ref{sec:masscons} involved only kinematic considerations, namely
mass conservation and the kinematic boundary conditions at the free surface
and bottom substrate.  Hence, everything we have done so far applies to any
incompressible flow.  Now we will restrict to the problem of a viscous
(Stokes) flow driven by gravity, which satisfies the equation
\begin{equation}
  \lapl\uv = \grad\pres - \gravv,
  \label{eq:Stokes}
\end{equation}
where~$\pres$ is the pressure and~$\gravv$ is a unit vector in the direction
of gravity.  The velocity satisfies the boundary conditions
\begin{subequations}
\begin{alignat}{2}
  \uv &= 0 \quad &\text{at } \yc = 0\,\qquad &\text{no-slip at substrate}\\
  \bm{t}_\alpha\cdot\stress\cdot\hat{\bm{n}} &= 0 \quad &\text{at } \yc = \eta
  \qquad
  &\text{tangential stresses at free surface}\\
  -\pres + \hat{\bm{n}}\cdot\stress\cdot\hat{\bm{n}} &= \sigma\kappasurf \quad
  &\text{at } \yc = \eta \qquad
  &\text{normal stress at free surface}\\
  \pd_\time\eta + \ucb^\alpha\,\pd_\alpha\eta &= \vcb
  &\text{at } \yc = \eta \qquad
  &\text{kinematic condition at free surface}
\end{alignat}\label{eq:BCs}%
\end{subequations}%
where
\begin{equation*}
  \stress \ldef \grad\uv + (\grad\uv)^T
\end{equation*}
is the deviatoric stress,~$\hat{\bm{n}}$ is the unit normal to the
surface,~$\bm{t}_\alpha$ are tangents to the surface, and~$\kappasurf$ is the
mean curvature of the surface.  The stress conditions assumes that the gas
above the surface has negligible viscosity and density.

We have absorbed the constant density, viscosity, and strength of gravity into
the velocity and pressure, so~$\uv$ stands for~$(\visc\uv/\grav)$,~$\pres$
stands for~$(\pres/\grav\rho)$, and the surface tension~$\sigma$ (introduced
below) stands for~$(\sigma/\grav\rho)$.  This means that the velocity
in~\eqref{eq:Stokes} has units of squared length, the pressure has units of
length, and the surface tension has units of squared length.

In the coordinates of Section~\ref{sec:coords}, the location of the surface is
\begin{equation*}
  \bar\rv(\xc^1,\xc^2,\time)
  = \Xv(\xc^1,\xc^2) + \eta(\xc^1,\xc^2,\time)\,\euv{3}(\xc^1,\xc^2)
\end{equation*}
and the surface tangents~$\bar{\bm{t}}_\alpha$ are then given by
\begin{equation}
  \bar{\bm{t}}_\alpha = \pd_\alpha\bar\rv
  = \evb{\alpha} + \pd_\alpha\eta\,\euv{3}.
  \label{eq:surftangents}
\end{equation}
The unit normal~$\hat{\bm{n}}$ is obtained by taking the cross product of the
tangents and normalizing.

In the coordinates introduced in Section~\ref{sec:coords}, with the velocity
field~\eqref{eq:vel}, The Stokes Eq.~\eqref{eq:Stokes} becomes
\begin{equation}
  \frac{1}{\wmgt}\l\{\pd_\alpha
  \l(\wmgt\,\widetilde\Metricc^{\alpha\beta}\pd_\beta\r)
  + \frac{\pd}{\pd\yc}\l(\wmgt\,\frac{\pd}{\pd\yc}\r)\r\}
  (\uct^\gamma\,\evt{\gamma} + \vct\,\euv{3})
  = \evtup{\alpha}\,\pd_\alpha\pres
  + \euv{3}\,\frac{\pd\pres}{\pd\yc} - \gravsc^\alpha\,\ev{\alpha}
  - \gravy\,\euv{3}\,,
  \label{eq:Stokescov}
\end{equation}
where we wrote
\begin{equation*}
  \gravv = \gravsc^\alpha\,\ev{\alpha} + \gravy\,\euv{3}
\end{equation*}
with components given by~\eqref{eq:gravvec}.  Note that, unlike the velocity
field~\eqref{eq:vel}, we have \emph{not} written the gravity vector in terms
of the bulk (tilde) coordinate vectors.  This is to avoid~$\yc$ dependence
in~$\gravsc^\alpha$ and~$\gravy$.

\subsection{Small-parameter Rescaling}
\label{sec:small}

Following RRS,~\cite{Roy2002} we rescale the variables to account for the
thinness of the fluid layer.  We do this not by rescaling~$\yc$, but by
rescaling~$\xc^\alpha$: we treat the variations in the substrate as being
large-scale, whilst the depth of the layer is of order unity.  Hence, we
introduce the rescalings
\begin{equation*}
  \xc^\alpha = \eps^{-1}\,{\xc^\alpha}^*,\quad
  \vct = \eps\,{\vct}^*,\quad
  \pres = \eps^{-1}\,{\pres}^*,\quad
  \sigma = \eps^{-2}\,\sigma^*,
\end{equation*}
where~$\eps$ is a small parameter; all other variables are left unscaled.
Following standard practice, we immediately drop the~${}^*$ superscripts and
deal only with the rescaled variables.  The amplitude of the variations of the
substrate are taken to be such that the vectors~$\ev{\alpha}$ and hence the
substrate metric~$\Metricc$ are of order unity.  All terms involving the
curvature tensor~$\mathbb{K}$ are of order~$\eps$, because of the
$\pd/\pd\xc^\alpha$ in the definition~\eqref{eq:curv}.

The continuity equation~\eqref{eq:divucov} is unchanged by the rescalings, and
neither is the kinematic boundary condition~\eqref{eq:kinematicBC}.  The
dynamical equation~\eqref{eq:Stokescov} becomes
\begin{equation}
  \frac{1}{\wmgt}\l\{\pow{\eps}{2}\,\pd_\alpha
  \l(\wmgt\,\widetilde\Metricc^{\alpha\beta}\pd_\beta\r)
  + \frac{\pd}{\pd\yc}\l(\wmgt\,\frac{\pd}{\pd\yc}\r)\r\}
  (\uct^\gamma\,\evt{\gamma} + \eps\,\vct\,\euv{3})
  = \,\evtup{\alpha}\,\pd_\alpha\pres
  + \eps^{-1}\euv{3}\,\frac{\pd\pres}{\pd\yc}
  - \gravsc^\alpha\,\ev{\alpha} - \gravy\,\euv{3}\,,
  \label{eq:Stokescoveps}
\end{equation}
where~$\wmgt = \wmg(1 - \eps\,\kappa\,\yc) +
\Order{\pow{\eps}{2}}$,~$\evt{\alpha} = \ev{\alpha} - \eps\,\yc\,
{\mathbb{K}_\alpha}^\beta\,\ev{\beta}$, and~$\evtup{\alpha} = \evup{\alpha} +
\eps\,\yc\, \mathbb{K}_{\alpha\beta}\,\evup{\beta} + \Order{\pow{\eps}{2}}$.
As for the boundary conditions~\eqref{eq:BCs}, they simplify to
\begin{subequations}
\begin{alignat}{2}
  \uct^\alpha
  &= \vct = 0 \quad &\text{at } \yc = 0\,;\\
  \frac{\pd\uct^\alpha}{\pd\yc} &= 0 + \Order{\pow{\eps}{2}} \quad
  &\text{at } \yc = \eta\,;\\
  \pres &= -\sigma\,\kappasurf
  + \Order{\pow{\eps}{2}} \quad &\text{at } \yc = \eta\,,
\end{alignat}\label{eq:BCseps}%
\end{subequations}%
where
\begin{equation*}
  \kappasurf = \kappa + \eps\,\eta\,\kappa_2 + \eps\lapl\eta +
  \Order{\pow{\eps}{2}}
\end{equation*}
with~$\lapl\eta=\grad_\alpha\grad^\alpha\eta =
\wmg^{-1}\pd_\alpha(\wmg\,\Metricc^{\alpha\beta}\pd_\beta\eta)$ the covariant
Laplacian of~$\eta$, and
\begin{equation*}
  \kappa_2 \ldef {\mathbb{K}_\alpha}^\beta\,{\mathbb{K}_\beta}^\alpha\,.
\end{equation*}
We shall not need the higher-order terms in~\eqref{eq:BCseps}.

\subsection{Velocity Field}
\label{sec:vel}

We now proceed in the usual manner and expand the fields as power series
in~$\eps$:
\begin{equation*}
\begin{split}
  \uct^\alpha &= \uct^\alpha_{(0)} + \eps\,\uct^\alpha_{(1)} + \ldots,\\
  \tilde\pres &= \tilde\pres_{(0)} + \eps\,\tilde\pres_{(1)} + \ldots,
\end{split}
\end{equation*}
where we have left out~$\vct$ since it is not needed in our development until
later (and it will not appear as an asymptotic series).  At order~$\eps^{0}$,
the velocity field and pressure satisfy
\begin{equation*}
  \frac{\pd^2\uct^{\alpha}_{(0)}}{\pd\pow{\yc}{2}} =
  \pd^\alpha\tilde\pres_{(0)} - \gravsc^\alpha\,,\qquad
  \frac{\pd\tilde\pres_{(0)}}{\pd\yc} = 0\,,
\end{equation*}
where~$\pd^\alpha = \Metricc^{\alpha\beta}\,\pd_\beta$.  These are readily
integrated to give
\begin{equation}
  \uct^{\alpha}_{(0)} = \AAA^\alpha_{(0)}\,\yc(\yc - 2\eta),\qquad
  \tilde\pres_{(0)} = -\sigma\kappa,
  \label{eq:uct0}
\end{equation}
where
\begin{equation}
  \AAA^\alpha_{(0)} = -\tfrac{1}{2}\l(\gravsc^\alpha
  + \sigma\,\pd^\alpha\kappa\r)
  \label{eq:AAA0}
\end{equation}
and the boundary conditions~\eqref{eq:BCseps} have been applied.

At order~$\eps$, 
\begin{equation*}
  \frac{\pd^2\uct^{\alpha}_{(1)}}{\pd\pow{\yc}{2}} =
  \l(\kappa {\delta_\beta}^\alpha
  + 2\,{\mathbb{K}_\beta}^\alpha\r)\frac{\pd\uct^{\beta}_{(0)}}{\pd\yc}
  + \pd^\alpha\tilde\pres_{(1)}
  + 2\yc\,{\mathbb{K}_\beta}^\alpha\pd^\beta\tilde\pres_{(0)}
  - \yc\,\gravsc^\beta\,{\mathbb{K}_\beta}^\alpha\,,\qquad
  \frac{\pd\tilde\pres_{(1)}}{\pd\yc} = \gravy\,,
\end{equation*}
with solution
\begin{equation}
  \uct^{\alpha}_{(1)} = \AAA^\alpha_{(1)}\yc\l(\yc - 2\eta\r)
  + \BBB^\alpha_{(1)}\yc\l(\pow{\yc}{2} - 3\pow{\eta}{2}\r)
  \label{eq:uct1}
\end{equation}
where
\begin{subequations}
\begin{align}
  \AAA^\alpha_{(1)} &= \tfrac{1}{2}\l(\gravsc^\alpha\kappa\eta
  + 3\gravsc^\beta\,{\mathbb{K}_\beta}^\alpha\eta
  - \gravy\,\pd^\alpha\eta
  + \sigma\l\{\eta\l(\kappa\,\pd^\alpha\kappa
  + 2\,{\mathbb{K}_\beta}^\alpha\pd^\beta\kappa\r)
  - \pd^\alpha(\kappa_2\eta + \lapl\eta)\r\}\r),\\
  \BBB^\alpha_{(1)} &= -\tfrac{1}{6}
  \l(\gravsc^\beta + \sigma\,\pd^\beta\kappa\r)
  \l(\kappa\,{\delta_\beta}^\alpha + 4\,{\mathbb{K}_\beta}^\alpha\r).
\end{align}
\end{subequations}
To simplify these expressions, we made use of
\begin{equation*}
  \pd_\alpha\gravy = \pd_\alpha(\gravv\cdot\euv{3})
  = -\gravv\cdot{\mathbb{K}_\alpha}^{\beta}\ev{\beta}
  = -\gravsc^\beta\,\mathbb{K}_{\alpha\beta}\,,
\end{equation*}
since~$\gravv$ is a constant vector.

Now that we have~$\uct^\alpha$ to first order in~$\eps$, we can insert it in
the definition of the mass flux vector~\eqref{eq:flux},
\begin{equation}
  \Qcb^\alpha = \int_0^\eta\arft\,\uct^\alpha d\yc
  = \int_0^\eta(1-\eps\,\kappa\,\yc)\l(\uct^\alpha_{(0)}
 + \eps\,\uct^\alpha_{(1)}\r) d\yc + \Order{\pow{\eps}{2}},
 \label{eq:thefluxint}
\end{equation}
which gives
\begin{equation}
  \Qcb^\alpha = \Qcb^\alpha_{\mathrm{grav}} + \Qcb^\alpha_{\mathrm{surf}}
  \label{eq:theflux}
\end{equation}
with
\begin{multline}
  \Qcb^\alpha_{\mathrm{grav}} = \tfrac{1}{3}\pow{\eta}{3}\l\{\gravsc^\alpha
  - \eps\,\gravsc^\beta\l(\kappa\,{\delta_\beta}^\alpha
  + \tfrac{1}{2}\,{\mathbb{K}_\beta}^\alpha\r)\eta
  + \eps\,\gravy\,\pd^\alpha\eta\r\}\\
  + \pow{\eps}{2}\tfrac{1}{120}\,\pow{\eta}{4}\kappa\,
  \{\eta\,\gravsc^\beta\,(9\kappa\,{\delta_\beta}^\alpha
    + 11\,{\mathbb{K}_\beta}^\alpha)
  - 25\,\gravy\,\pd^\alpha\eta\} + \Order{\pow{\eps}{2}},
  \label{eq:thefluxgrav}
\end{multline}
\begin{multline}
  \Qcb^\alpha_{\mathrm{surf}} = 
  \tfrac{1}{3}\sigma\pow{\eta}{3}\l\{\pd^\alpha\kappasurf
  - \eps\,\eta\,\kappa\,\pd^\alpha\kappa
  + \tfrac{1}{2}\eps\,\eta\,{\mathbb{K}_\beta}^\alpha\,\pd^\beta\kappa\r\}\\
  + \pow{\eps}{2}\tfrac{1}{120}\,\sigma\,\pow{\eta}{4}\kappa\,
  \{9\eta\,\kappa\,\pd^\alpha\kappa
  - 14\,\eta{\mathbb{K}_\beta}^\alpha\pd^\beta\kappa
  - 25\,\pd^\alpha\!\l(\kappa_2\eta + \lapl\eta\r)\} + \Order{\pow{\eps}{2}},
  \label{eq:thefluxsurf}
\end{multline}
where we have kept some second-order terms but neglected others: this will be
explained when we find~$\vct$ below and is related to the exact preservation
of the kinematic constraints of the problem.  Equation~\eqref{eq:masscons}
with the flux~\eqref{eq:theflux} is the dynamical equation that governs the
evolution of~$\eta$.  To order~$\eps$, it is the same equation as that found
by RRS,~\cite{Roy2002} but in nonorthogonal coordinates rather than orthogonal
coordinates.  This is a straightforward generalization, but it is illuminating
to rederive the equations in this setting, and allows a thorough introduction
to all the notation.  We will also need the quadratic terms
in~\eqref{eq:theflux} below, and these are not present in RRS's formulation:
this is not a failure of RRS but rather is required by the specific
application we have in mind, namely particle advection.

In order to do particle advection in Section~\ref{sec:advection}, we will need
the vertical velocity field,~$\vct$.  It is obtained by integrating the
continuity equation~\eqref{eq:divucov},
\begin{equation}
  \vct
  = -\frac{1}{\wmgt}\int_0^\yc\pd_\alpha(\wmgt\,\uct^\alpha)\,d\yc
  \label{eq:vc}
\end{equation}
where the integral's lower bound enforces the no throughflow boundary
condition at~$\yc=0$.  The kinematic boundary condition~\eqref{eq:kinematicBC}
is satisfied by~\eqref{eq:vc} at~$\yc=\eta$, since
\begin{align}
  \vcb = -\frac{1}{\wmgb}\int_0^\eta\pd_\alpha(\wmgt\,\uct^\alpha)\,d\yc
  &= -\frac{1}{\wmgb}\,\pd_\alpha\int_0^\eta(\wmgt\,\uct^\alpha)\,d\yc
  + \frac{1}{\wmgb}\,(\wmgb\,\ucb^\alpha)\pd_\alpha\eta\nonumber\\
  &= -\frac{1}{\wmgb}\,\pd_\alpha(\wmg\,\Qcb^\alpha)
  + \ucb^\alpha\,\pd_\alpha\eta\,.
  \label{eq:kbv}
\end{align}
The first term on the right-hand side of~\eqref{eq:kbv} vanishes because
of~\eqref{eq:masscons}, and the second gives the kinematic boundary
condition~\eqref{eq:kinematicBC}.  But observe that to recover this boundary
condition we need factors of~$\wmgb$ to cancel; hence, a factor of~$\wmgb$
must appear multiplicatively in the integrand if we are to satisfy the
kinematic boundary condition \emph{exactly} (\ie, to all orders in~$\eps$).
This is the reason for keeping some second-order terms
in~\eqref{eq:thefluxint} and~\eqref{eq:theflux}.  Note that this way the
divergence-free condition~\eqref{eq:divucov} is also satisfied exactly.

The easiest way to get the~$\yc$ coefficients for~$\vc$ is to
combine~\eqref{eq:uct0} and~\eqref{eq:uct1} into
\begin{equation}
  \uct^{\alpha} = \uct^{\alpha}_{(0)} + \eps\,\uct^{\alpha}_{(1)}
  = \AAA^\alpha\yc\l(\yc - 2\eta\r)
  + \BBB^\alpha\yc\l(\pow{\yc}{2} - 3\pow{\eta}{2}\r),
  \label{eq:uct}
\end{equation}
valid to first order in~$\eps$, where~$\AAA^\alpha = \AAA^\alpha_{(0)} +
\eps\AAA^\alpha_{(1)}$ and~$\BBB^\alpha = \eps\BBB^\alpha_{(1)}$.  Then,
using~$\wmgt=\wmg(1-\eps\kappa\yc)$, we find from~\eqref{eq:vc}
\begin{equation}
  \vct = \frac{\pow{\yc}{2}}{1-\eps\kappa\yc}
  \l(\aaa + \bbb\,\yc + \ccc\,\pow{\yc}{2} + \ddd\,\pow{\yc}{3}\r)
  \label{eq:vct}
\end{equation}
where
\begin{alignat*}{2}
  \aaa &= \tfrac{1}{2}\grad_\alpha\l(2\eta\AAA^\alpha +
  3\pow{\eta}{2}\BBB^\alpha\r),\qquad
  &\ccc &= \tfrac{1}{4}\grad_\alpha\l(\eps\kappa\AAA^\alpha - \BBB^\alpha\r),\\
  \bbb &= -\tfrac{1}{3}\grad_\alpha\l((1+2\eps\kappa\eta)\AAA^\alpha
  + 3\eps\kappa\pow{\eta}{2}\BBB^\alpha\r),\qquad
  &\ddd &= \tfrac{1}{5}\grad_\alpha\l(\eps\kappa\BBB^\alpha\r),
\end{alignat*}
and the covariant divergence is defined after Eq.~\eqref{eq:almosthere}.  The
expression~\eqref{eq:vct} for~$\vct$ is exact, given the first-order
expressions~$\wmgt=\wmg(1-\eps\kappa\yc)$ and~\eqref{eq:uct}
for~$\uct^\alpha$.

\section{Solution of the Equations}
\label{sec:PDEsolve}

We now seek steady solutions of Eq.~\eqref{eq:masscons} with the mass flux
vector~\eqref{eq:theflux}.  The unknown function is the thickness
field~$\eta(\xc^1,\xc^2)$.  We characterize the shape of the substrate using
the parametrization in Section~\ref{sec:monge}, and assume the substrate is
periodic in~$\xc^1$ and~$\xc^2$ with unit period. \savecomment{This is maybe
too strong.  Also, I still haven't modified the code to allow for different
periodicities.}  In dimensionless form, the independent input parameters of
the problem are thus
\begin{enumerate}
  \item The inclination angle~$\theta$ and rotation angle~$\phi$ of the
  gravity vector [Eq.~\eqref{eq:gravvec}];\label{item:angles}
  \item The thickness of the film at the
  corner,~$\eta(0,0)=\eps$;\label{item:eps}
  \item The shape of the substrate, including its height, specified
  with~$\fmg(\xc^1,\xc^2)$;\label{item:shape}
  \item The surface tension~$\sigma$.
\end{enumerate}
The parameters~\ref{item:angles} determines the speed and direction of the
flow: the fluid is static for~$\theta=0$.  Parameter~\ref{item:eps}, the
thickness of the film at the corner, effectively determines~$\eps$.  We shall
thus drop~$\eps$ from the equations for the remainder of the paper, instead
introducing the small parameter through~$\eta(0,0)=\eps$.
Parameter~\ref{item:shape} is infinitely rich: even for a given substrate
shape, we can still vary its height by scaling~$\fmg$ by a constant.  We shall
limit ourselves to substrates of relatively simple harmonic shapes.

\subsection{Substrates with Translational Symmetry}
\label{sec:transsub}

Consider first the case where the substrate has a translational symmetry.  We
may then orient our coordinates such that~$\fmg$ is only a function
of~$\xc^2$.  From Eq.~\eqref{eq:mongeparams}, we then have~$\pmg=\smg=\tmg=0$.
Since nothing depends on~$\xc^1$, we seek solutions of the form~$\eta(\xc^2)$.
Equation~\eqref{eq:masscons} then reduces to
\begin{equation}
  \pd_2(\wmg\,\Qcb^2) = 0 \quad\Longrightarrow\quad
  \wmg\,\Qcb^2 = \text{const.}
  \label{eq:1Deq}
\end{equation}
since~$\pd_1(\wmgt\Qcb^1)$ vanishes.  If we keep only the~$\eps^0$ terms
in~\eqref{eq:theflux}, we can solve~\eqref{eq:1Deq} analytically,
\begin{equation}
  \eta(\xc^2) = \l(c\,\wmg\,(\gravsc^2
  + \sigma\,\wmg^{-2}\,\pd_2\kappa)\r)^{-1/3}
  \label{eq:1dsol}
\end{equation}
where the constant~$c$ is adjusted to satisfy the boundary
condition~$\eta(0)=\eps$, and we have used~$\Metricc^{22}=\wmg^{-2}$.  The
quantity in parentheses in~\eqref{eq:1dsol} can become singular or negative,
an indication that the higher-order terms can no longer be neglected.
\savecomment{Is this associated with ill-posedness of the equations?}  In that
case~$\pd_2(\wmg\Qcb^2) = 0$ still holds but we then have to solve an ODE
for~$\eta(\xc^2)$.  We shall not do this here and instead solve the general
PDE in the next section.  For comparison, the zeroth-order
solution~\eqref{eq:1dsol} is plotted along with the numerical first-order
solution for a sinusoidal substrate and corner thickness~$\eps=0.02$.
\begin{figure}
\psfrag{x1}{$\xc^1$}
\psfrag{x2}{$\xc^2$}
\psfrag{y}{$\yc$}
\centering
\includegraphics[width=.6\textwidth]{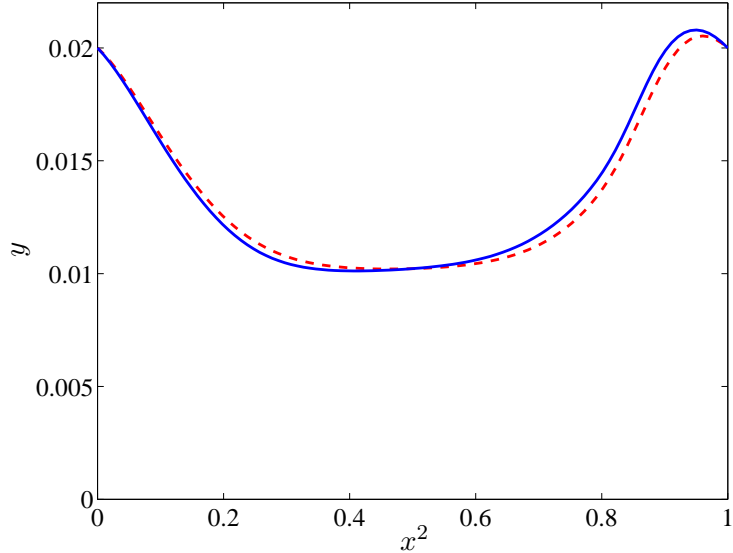}
\caption{Surface thickness~$\eta(\xc^2)$ for a substrate with translational
  symmetry~$\fmg(\xc^2)=0.1\,\sin(2\pi\xc^2)$,
  for~$\eps=0.02$,~$\theta=0.7$,~$\phi=1.2$, and~$\sigma=0$.  The solid
  line is the zeroth-order analytic solution~\eqref{eq:1dsol}, and the dashed
  line is the first-order numerical solution.}
\label{fig:1dsol}
\end{figure}
The two agree well since the thickness is small.

Note that even though the velocity field depends only on two variables
($\xc^2$ and~$\yc$), the flow is actually three-dimensional, since~$\AAA^1$
and~$\BBB^1$ in~\eqref{eq:uct} do not in general vanish, except in the special
case~$\phi=\pi/2$.  We shall see in Section~\ref{sec:advection} that even this
simple solution supports rather complicated trajectories.

\subsection{General Numerical Solution}
\label{sec:gensol}

In general, Eq.~\eqref{eq:masscons} with the mass flux
vector~\eqref{eq:theflux} must be solved numerically.  We use a fourth-order
finite-difference scheme together with Newton--Kantorovich
iteration~\cite{Boyd} to converge to the steady solution.  Since the N--K
scheme employs a linearization of the nonlinear equation to seek an improved
guess, it allows us to check the linear stability of the flow for free, and
all the flows presented here are stable.
\begin{figure}
\psfrag{x1}{\raisebox{-.4em}{$\xc^1$}}
\psfrag{x2}{\raisebox{.3em}{$\xc^2$}}
\centering
\subfigure[]{
  \includegraphics[width=.46\textwidth]{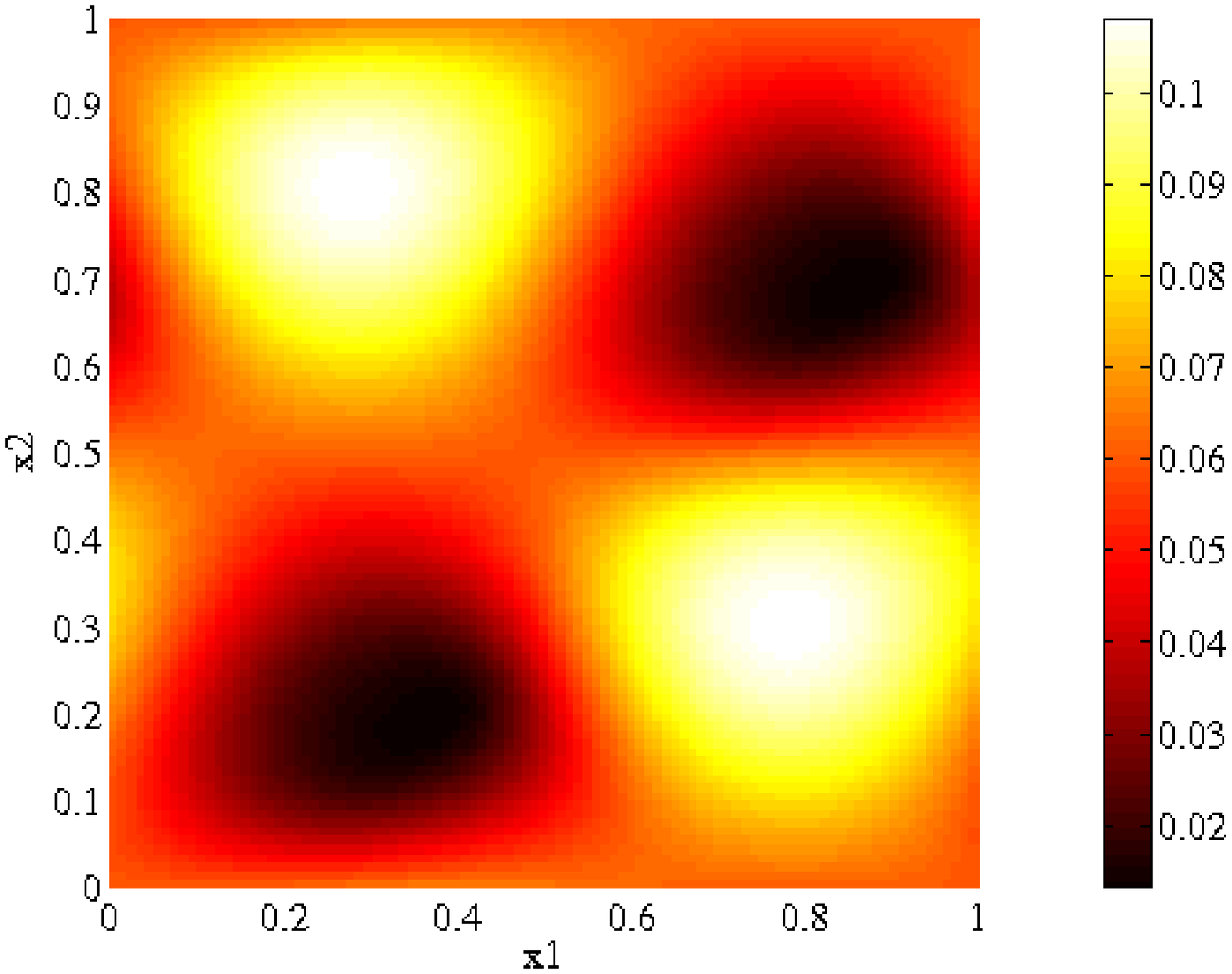}
  \label{fig:chaos_hot}
}\hspace{1em}%
\subfigure[]{
  \psfrag{x2}{$\xc^2$}
  \includegraphics[width=.46\textwidth]{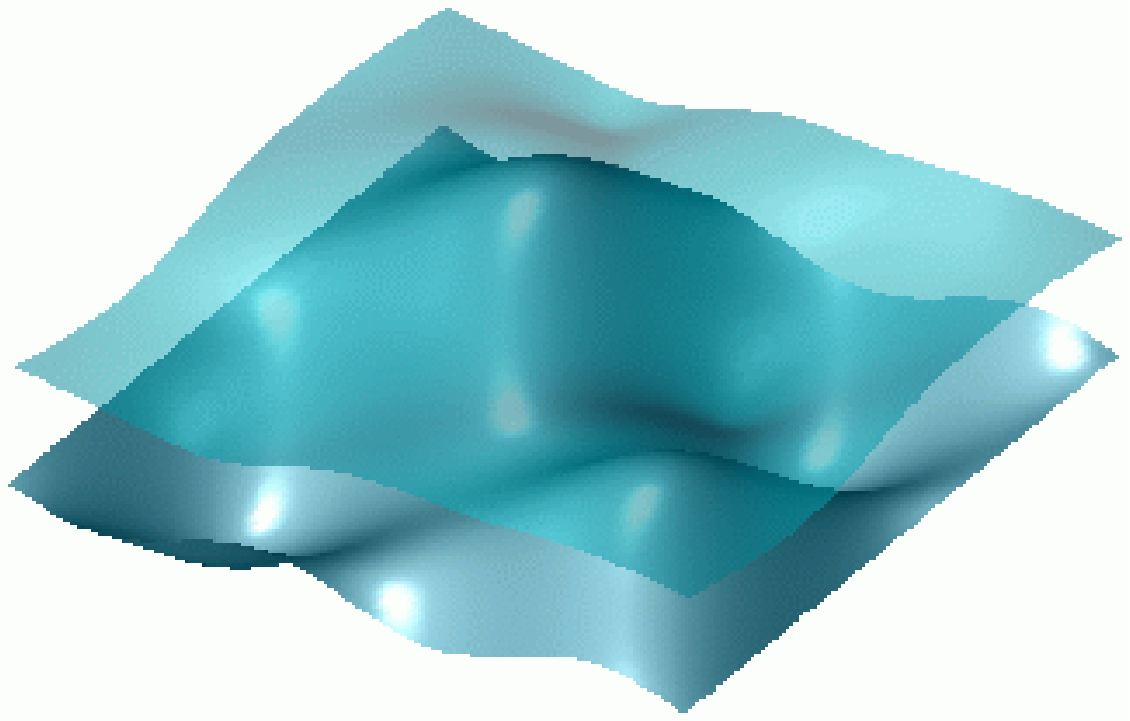}
  \label{fig:chaos_surf}
}
\caption{The thickness field~$\eta(\xc^1,\xc^2)$ for the substrate shape given
  by Eq.~\eqref{eq:chaos_subst}, with~$\eps=0.06$,~$\theta=0.1$,~$\phi=0$,
  and~$\sigma=0$.  (a) Contour plot: the flow is towards the right; (b)
  Surface plot: the flow is towards the upper-right.  The vertical scale is
  exaggerated.}
\label{fig:chaos_sol}
\end{figure}
In Figure~\ref{fig:chaos_sol} we show the thickness field of a typical
solution, for the substrate
\begin{equation}
  \fmg(\xc^1,\xc^2)=0.05\,(\sin(2\pi\xc^1)\sin(2\pi\xc^2) +
  0.2\sin(4\pi\xc^2))
  \label{eq:chaos_subst}
\end{equation}
with~$\eps=0.06$,~$\theta=0.1$,~$\phi=0$, and~$\sigma=0$.  As mentioned in the
introduction, we will set the surface tension to zero.  The solutions
presented in this paper typically have resolutions of~$100\times100$
gridpoints, which is well-resolved enough.  The acid test to see if we have
enough numerical resolution is whether particle orbits `spiral in' to
attractors, which is precluded if the flow is incompressible.  For example,
the orbits in the regular islands of Fig.~\ref{fig:psec_2D_nochaos} are nice
and closed over many periods.

When doing particle advection in Section~\ref{sec:advection} the thickness
field is first resampled on a grid of~$1024\times1024$ using Fourier
interpolation, and then the velocity field is evaluated off-gridpoints using
linear interpolation.  This has the advantage that the hardest part (the
resampling) only needs to be done once at the beginning and stored.  This is
cheaper, for instance, than using a higher-order interpolation scheme on the
coarser grid.

\section{Fluid Particle Trajectories}
\label{sec:advection}

To investigate the transport properties of the flow, we solve the advection
equation
\begin{equation}
  \pd_\time\rv(\time) = \uv(\rv(\time)), \qquad \rv(0) = \rv_0,
  \label{eq:advection}
\end{equation}
for a steady velocity field~$\uv$ of the form~\eqref{eq:vel} obtained
numerically as described in Section~\ref{sec:PDEsolve}.  Instead of converting
our velocity field back to Cartesian coordinates, we use the
definition~\eqref{eq:r} of~$\rv(\time)$ and take its time derivative, assuming
that the coordinates are a function of time, and use the chain rule:
\begin{equation*}
  \pd_\time\rv = \pd_\alpha\rv\,\pd_\time\xc^\alpha
  + \pd_\yc\,\rv\,\pd_\time\yc
  = \pd_\time\xc^\alpha\,\evt{\alpha} + \pd_\time\yc\,\euv{3}\,,
\end{equation*}
so that~\eqref{eq:advection} becomes \savecomment{This would change for
explicit time-dependence?}
\begin{equation}
  \pd_\time\xc^\alpha = \uct^\alpha,\qquad \pd_\time\yc = \vct\,.
  \label{eq:advectionc}
\end{equation}
We can thus follow particle trajectories in out curved coordinates
using~\eqref{eq:advectionc}.  We will now do so for different types of
substrates: those possessing translational symmetry
(Section~\ref{sec:transsym}) and those leading to chaotic trajectories
(Section~\ref{sec:chaos}).  In all cases, the particle trajectories are
computed using an adaptive embedded fourth/fifth-order Runge--Kutta
scheme~\cite{Dormand1980,Cash1990} and the velocity field is interpolated as
described in Section~\ref{sec:gensol}.

\subsection{Substrates with Translational Symmetry}
\label{sec:transsym}

\begin{figure}
\centering
\psfrag{x1=0}{\raisebox{.2em}{$\xc^1=0$}}
\psfrag{x2}{\raisebox{-.3em}{$\xc^2$}}
\psfrag{y}{\raisebox{.3em}{$\yc$}}
\subfigure[]{
  \includegraphics[width=.45\textwidth]{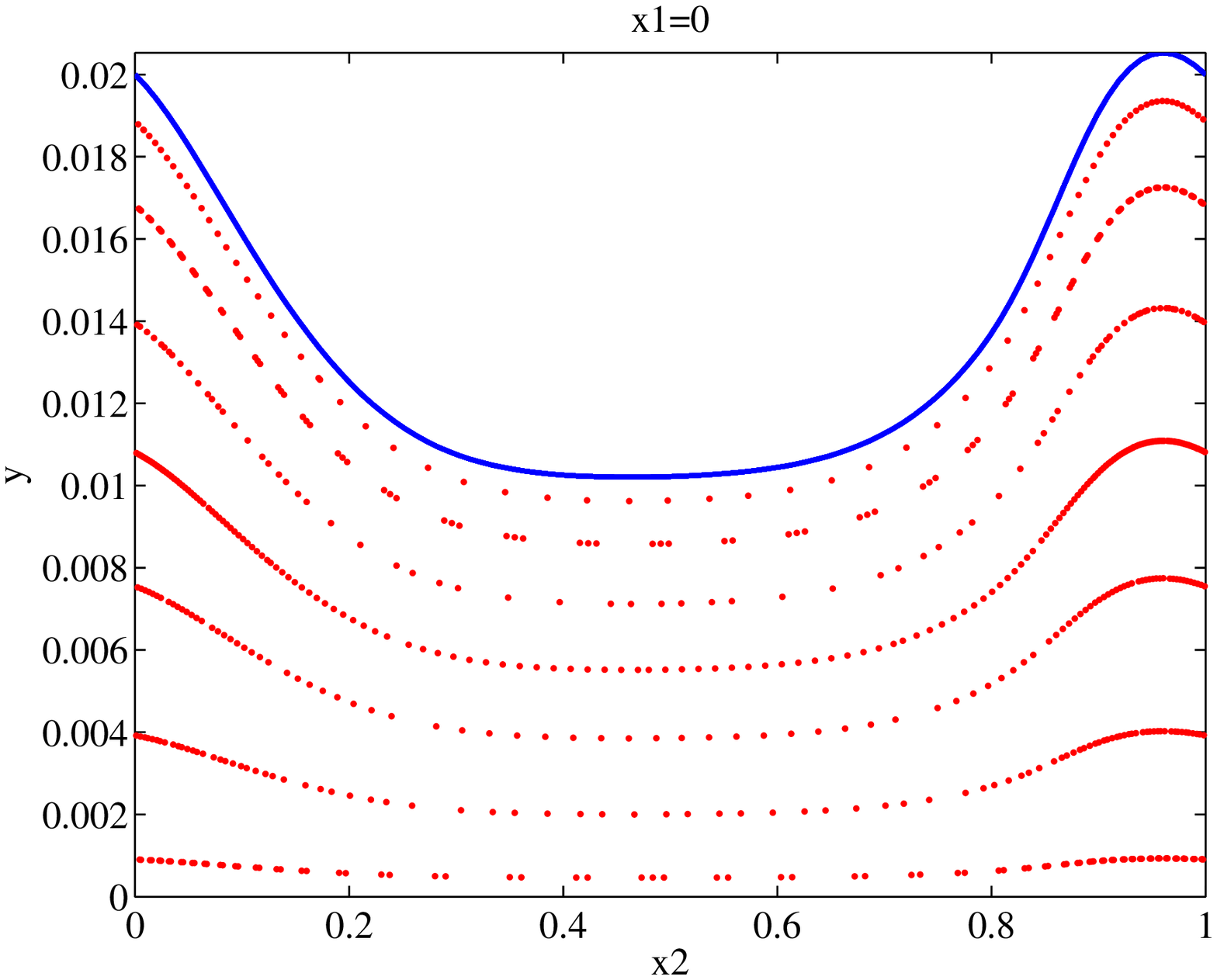}
  \label{fig:psec_1D_norolls}
}\goodgap%
\subfigure[]{
  \includegraphics[width=.45\textwidth]{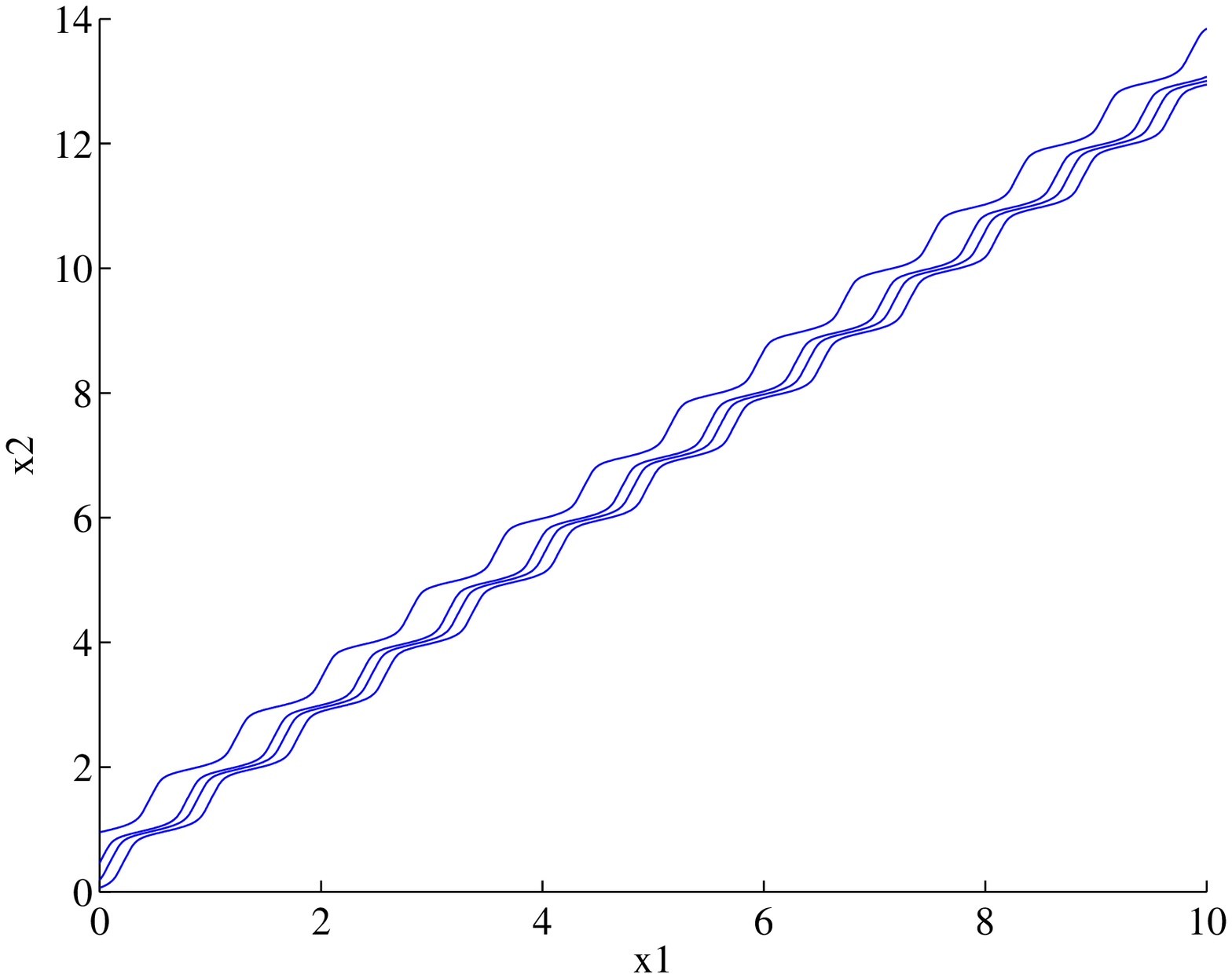}
  \label{fig:traj_1D_norolls}
}
\caption{For the same parameters as in Fig.~\ref{fig:1dsol}: (a) Poincar\'e
  section at~$\xc^1=0$; (b) Four typical trajectories at different
  depths~$\yc$.  The trajectories have the same overall slope.}
\label{fig:psec_1D}
\end{figure}

For substrates with translational symmetry, as described in
Section~\ref{sec:transsub}, the system~\eqref{eq:advectionc} is
\begin{equation}
  \pd_\time\xc^1 = \uct^1(\xc^2,\yc),\quad
  \pd_\time\xc^2 = \uct^2(\xc^2,\yc),\quad \pd_\time\yc = \vct(\xc^2,\yc).
  \label{eq:advectionc1D}
\end{equation}
We may immediately write~$\xc^1 = \uct^1(\xc^2,\yc)\,\time + \xc^1_0$, so
that~\eqref{eq:advectionc1D} is effectively an autonomous two-dimensional
system, which is solvable by quadratures and cannot exhibit
chaos.~\cite{Eckmann1985} The system admits a streamfunction~$\psi(\xc^2,\yc)$
such that
\begin{equation*}
  \uct^2 = \wmgt^{-1}\,\pd_\yc\psi, \qquad
  \vct = -\wmgt^{-1}\,\pd_2\psi
\end{equation*}
from which it is easily seen that~$\psi = \wmg\,\Qct^2$, defined
in~\eqref{eq:flux}.  If the zeroth-order terms dominate everywhere, the
resulting velocity profile is quadratic in~$\yc$ (`half-Poiseuille') and the
flow in the~$\xc^2$--$\yc$ plane cannot have a stagnation point inside the
fluid, except at~$\yc = 0$.  Thus, a Poincar\'e section of trajectories, made
by recording intersections with the plane~$\xc^1=0$, must look like
Fig.~\ref{fig:psec_1D_norolls}, where every trajectory crosses the domain
in~$\xc^2$ by going over furrows and crests.
In Fig.~\ref{fig:traj_1D_norolls} we see that four typical trajectories have
roughly the same overall `slope.'  This leads to poor transport properties,
since particles initially near each other remain near.  This is due to the
thinness of the layer which causes the zeroth-order velocity to dominate.  If
we use the zeroth order horizontal velocity~\eqref{eq:uct0}, we have
\begin{equation*}
  \uct^{2}_{(0)}/\uct^{1}_{(0)} = \AAA^2_{(0)}/\AAA^1_{(0)}
\end{equation*}
independent of the depth~$\yc$.  This implies that when the layer is very thin
the trajectories in Fig.~\ref{fig:traj_1D_norolls} are independent of depth.
(This is true for any substrate, not just those with translational symmetry.)
\savecomment{This argument could be made better: rule out interesting dynamics
unless first-order terms are included.}  First-order corrections terms are
required in order to have more complicated dynamics.

\begin{figure}
\psfrag{x1=0}{\raisebox{.2em}{$\xc^1=0$}}
\psfrag{x1}{\raisebox{-.3em}{$\xc^1$}}
\psfrag{y}{\raisebox{.3em}{$\yc$}}
\centering
\subfigure[]{
  \psfrag{x2}{\raisebox{-.3em}{$\xc^2$}}
  \includegraphics[width=.48\textwidth]{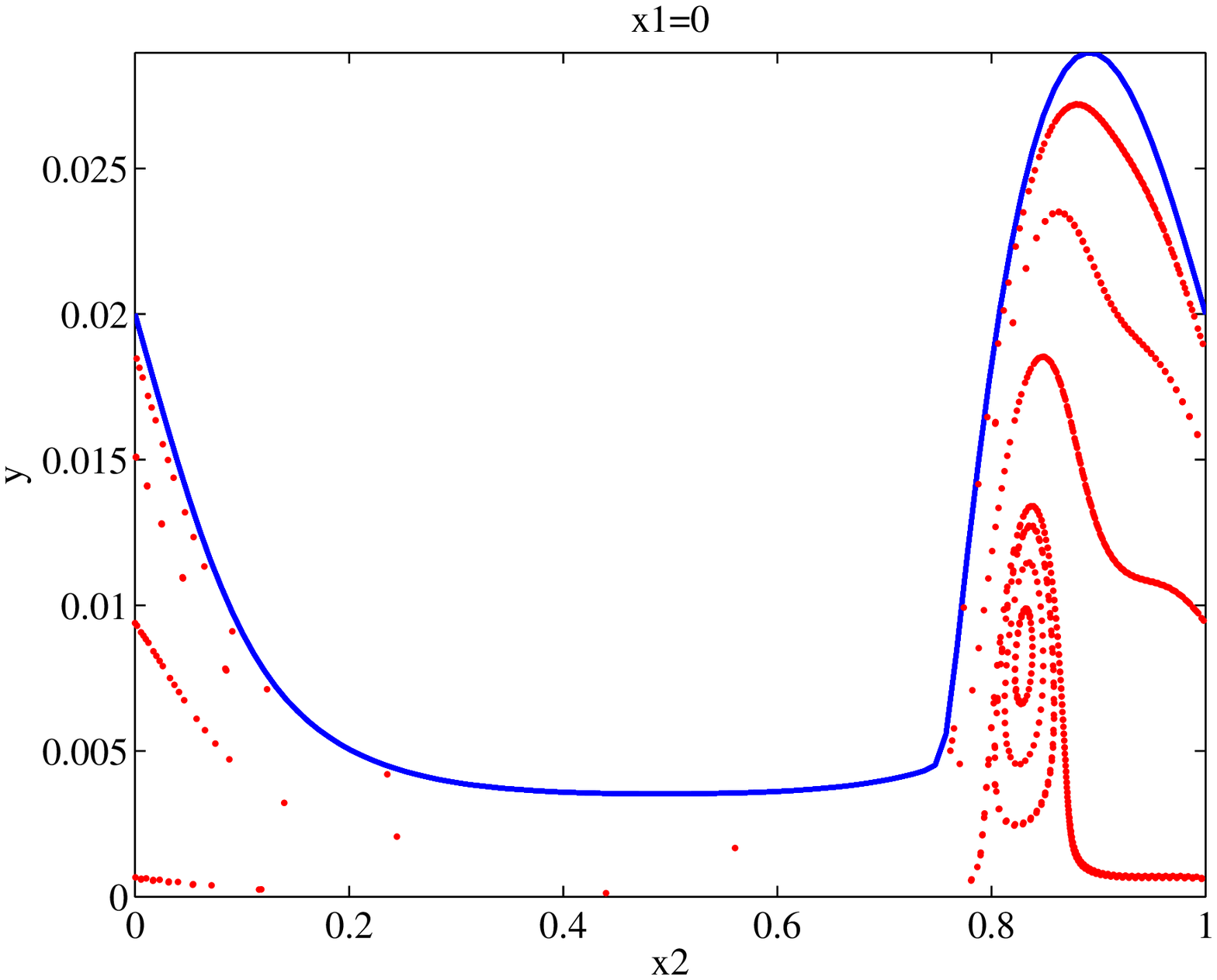}
  \label{fig:psec_1D_rolls}
}\hspace{0em}%
\subfigure[]{
  \psfrag{x2}{$\xc^2$}
  \includegraphics[width=.48\textwidth]{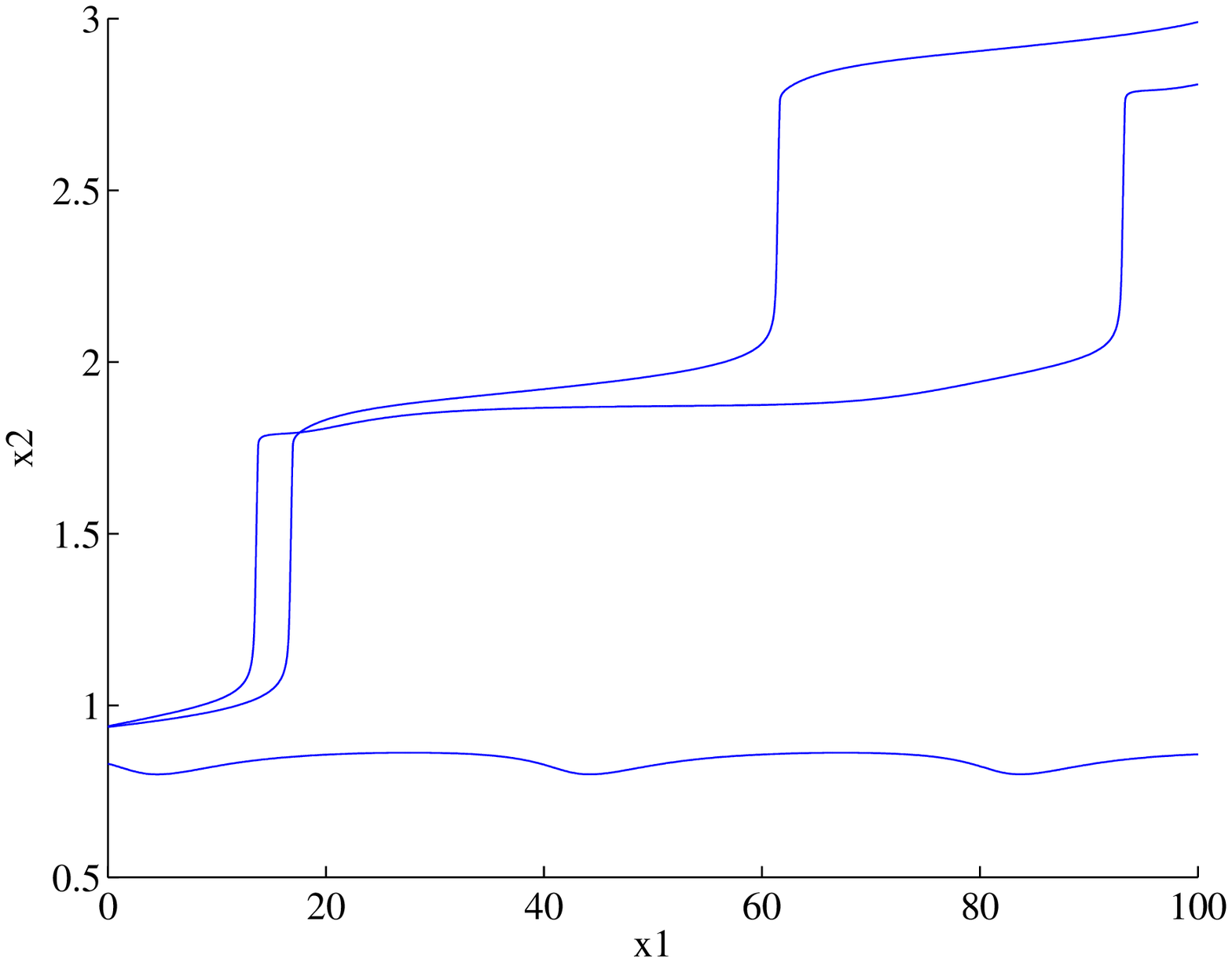}
  \label{fig:traj_1D_rolls}
}
\caption{For the same parameters as in Fig.~\ref{fig:1dsol} but
  with~$\theta=0.6$, $\phi=0.8$: (a) Poincar\'e section at~$\xc^1=0$, showing
  no apparent chaos; (b) The types of trajectories seen in (a): the bottom
  trajectory is trapped in a furrow, whilst the top two trajectories moves
  rapidly over crests, and more slowly over furrows.  The three trajectories
  start at different depths~$\yc$.  Four typical trajectories, showing a
  trapped trajectory and three trajectories that undergo regular
  `near-trapping' events.  Note that the overall slopes of the trajectories
  are different, and are not monotonic with depth.}
\label{fig:1D_rolls}
\end{figure}

If the inclination angle~$\theta$ of the substrate is decreased, or~$\phi$ is
changed so that the furrows present a more acute angle to the incoming flow,
then a situation as in Fig.~\ref{fig:1D_rolls} develops.  Here some of the
fluid cannot make it over the crests and is trapped in a recirculating region.
The trajectories that are not trapped go very quickly when over crests, as
indicated by the paucity of points in the shallow regions of the Poincar\'e
section Fig.~\ref{fig:psec_1D_rolls}, and linger over furrows, where the fluid
is deeper.  This is of course a consequence of mass conservation, as the fluid
is forced to flow more rapidly in the shallows.  Three representative
trajectories are depicted in Figure~\ref{fig:traj_1D_rolls}, where the change
in speed in the untrapped trajectory (top) is reflected in the slope of the
curve.
In contrast to Fig.~\ref{fig:traj_1D_norolls}, the overall slope of a
trajectory now depends on its depth.  This important effect,
reflected in the different slopes of the trajectories in
Fig.~\ref{fig:traj_1D_rolls}, leads to enhanced transport, in the sense that
particles at different depths have very different histories.
Furthermore, the slope (or lateral drift rate) of a trajectory is not
monotonic in the depth, since it is due to a complicated resonance between
the lateral motion and downslope topography.

If the inclination angle~$\theta$ is too small, or~$\phi$ is such that the
furrows are too aligned with the incoming flow, then untrapped trajectories
disappear but the flow also develops dry spots, since the fluid doesn't make
it over the crests of the bumps: it is flowing in the `gutters' of the
substrate.  This takes us outside the scope of the present theory and we do
not consider such cases.

\subsection{Chaotic Trajectories}
\label{sec:chaos}

In Section~\ref{sec:transsym} we looked at trajectories for the simplest type
of substrate, those possessing a translational symmetry.  We now want to
investigate the potential for creating flows that exhibit chaotic
trajectories, leading to chaotic advection.~\cite{Aref1984,Ottino,Wiggins2004}
Breaking the translational symmetry of the substrate is essential for the
existence of chaotic trajectories, otherwise the trajectories obey a
two-dimensional autonomous system, as in Section~\ref{sec:transsym}, which
cannot exhibit chaos.  Given that this translational symmetry is broken, there
are three modifications to the flow that increase the chances of
developing chaotic trajectories:
\begin{enumerate}
\item Break discrete symmetries of the substrate;
\item Make the fluid deeper (but still thin);
\item Make the inclination angle~$\theta$ smaller.
\end{enumerate}
The first point removes barriers to transport associated with
symmetries.~\cite{Franjione1989,%
Franjione1992,Ottino1992,Jana1994,Mezic1994}
The second point allows particles more vertical freedom to break the
two-dimensional constraint that leads to integrable motion, by bringing in
first-order terms in the velocity field.  The third point
allows the particles more time to move laterally during one periodic length of
the domain by decreasing the flow velocity.  Here, by laterally we mean a
direction perpendicular to the mean flow.

As an illustration we will use the substrate with height function given
by~\eqref{eq:chaos_subst}.  The first term in~\eqref{eq:chaos_subst} breaks
the translational symmetry, and the second term is a perturbation that breaks
the discrete symmetry~$\xc^2 \rightarrow \xc^2 + \tfrac{1}{2}$.  Thus, the
substrate with those parameter values has a good chance of exhibiting chaos,
according to our first point above.

\begin{figure}
\psfrag{x1=0}{\raisebox{.2em}{$\xc^1=0$}}
\psfrag{x1}{\raisebox{-.3em}{$\xc^1$}}
\psfrag{y}{\raisebox{.3em}{$\yc$}}
\centering
\subfigure[]{
  \psfrag{x2}{\raisebox{-.3em}{$\xc^2$}}
  \includegraphics[width=.48\textwidth]{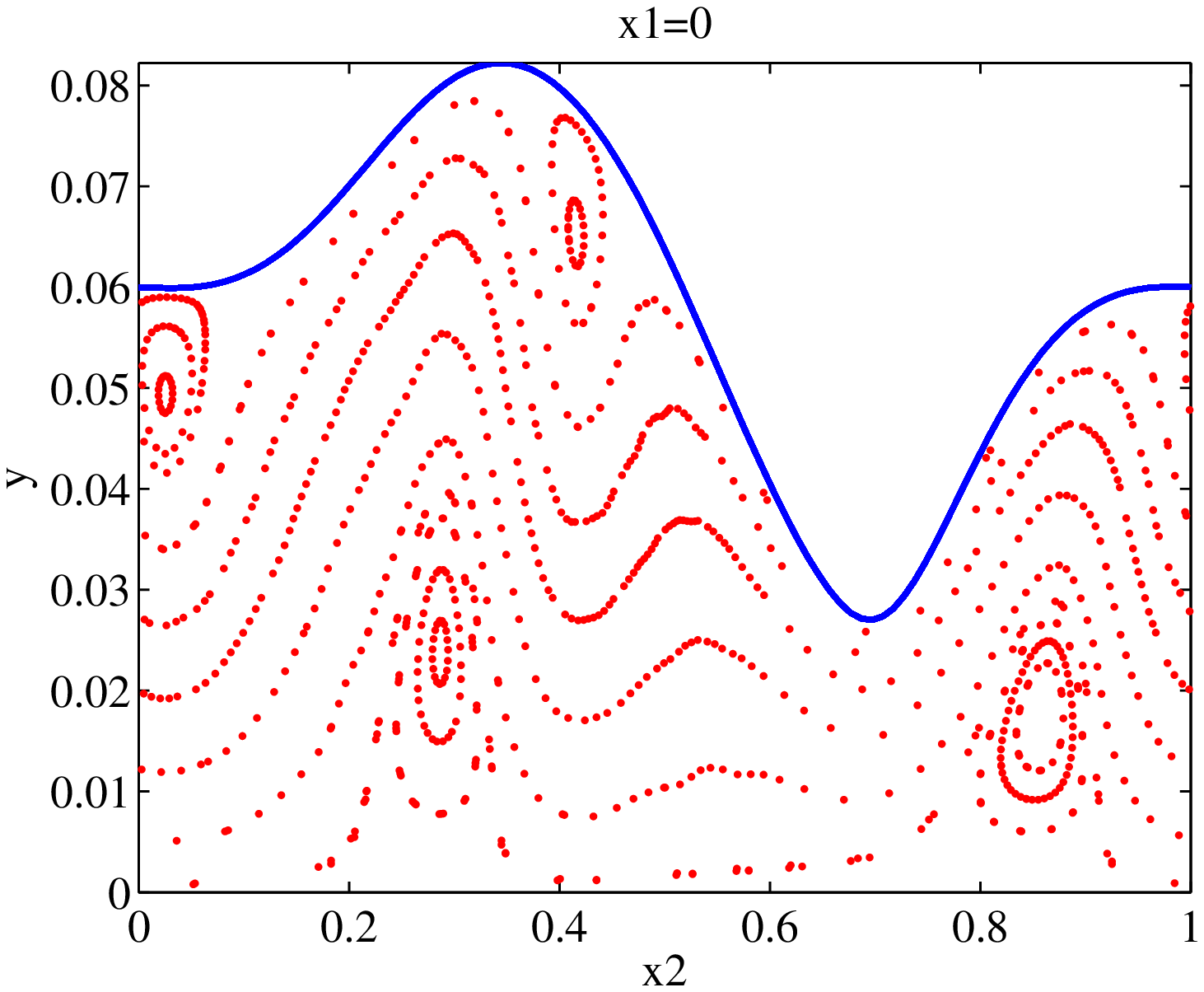}
  \label{fig:psec_2D_nochaos}
}\hspace{0em}%
\subfigure[]{
  \psfrag{x2}{$\xc^2$}
  \includegraphics[width=.48\textwidth]{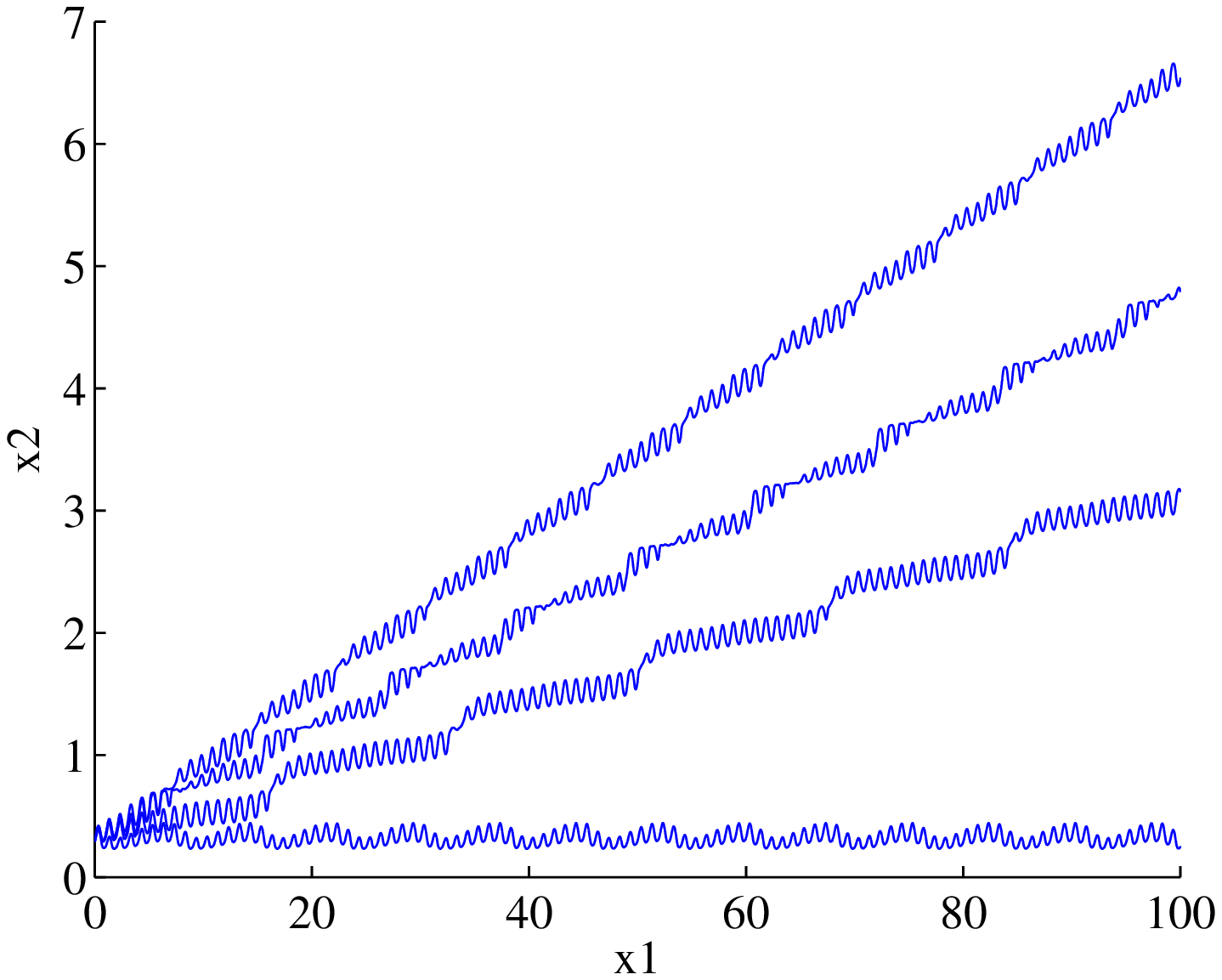}
  \label{fig:traj_2D_nochaos}
}
\caption{For the substrate shape~\eqref{eq:chaos_subst} with
  corner-thickness~$\eps=0.06$ and inclination angle~$\theta=0.1$: (a)
  Poincar\'e section at~$\xc^1=0$, showing no apparent chaos; (b) Four typical
  trajectories, showing a trapped trajectory and three trajectories that
  undergo regular `near-trapping' events.  Note that the overall slopes of the
  trajectories are different, and are not monotonic with depth.}
\label{fig:nochaos}
\end{figure}

Consider for an inclination angle~$\theta=0.2$ for the
substrate~\eqref{eq:chaos_subst}, with a corner thickness~$\eps=0.06$.  This
means that the fluid is reasonably thick, so our second point is
satisfied.  But the Poincar\'e section in Fig.~\ref{fig:psec_2D_nochaos}
clearly shows that no chaos is present, since the phase space is foliated by
one-dimensional curves.  (Though very small regions of chaos could in
principle be present, they are clearly not important here.)
Figure~\ref{fig:traj_2D_nochaos} shows four typical trajectories, taken at
about the same initial~$\xc^2$ but at different depths~$\yc$.  The three
untrapped trajectories undergo `near-trapping,' in the sense that they appear
to flirt with the idea of entering the trapped regions.  This motion is
periodic, but the period depends on the trajectory.  There is thus a
depth dependence as for the case in Fig.~\ref{fig:1D_rolls}.

\begin{figure}
\psfrag{x1=0}{\raisebox{.2em}{$\xc^1=0$}}
\psfrag{x1}{\raisebox{-.3em}{$\xc^1$}}
\psfrag{y}{\raisebox{.3em}{$\yc$}}
\centering
\subfigure[]{
  \psfrag{x2}{\raisebox{-.3em}{$\xc^2$}}
  \includegraphics[width=.48\textwidth]{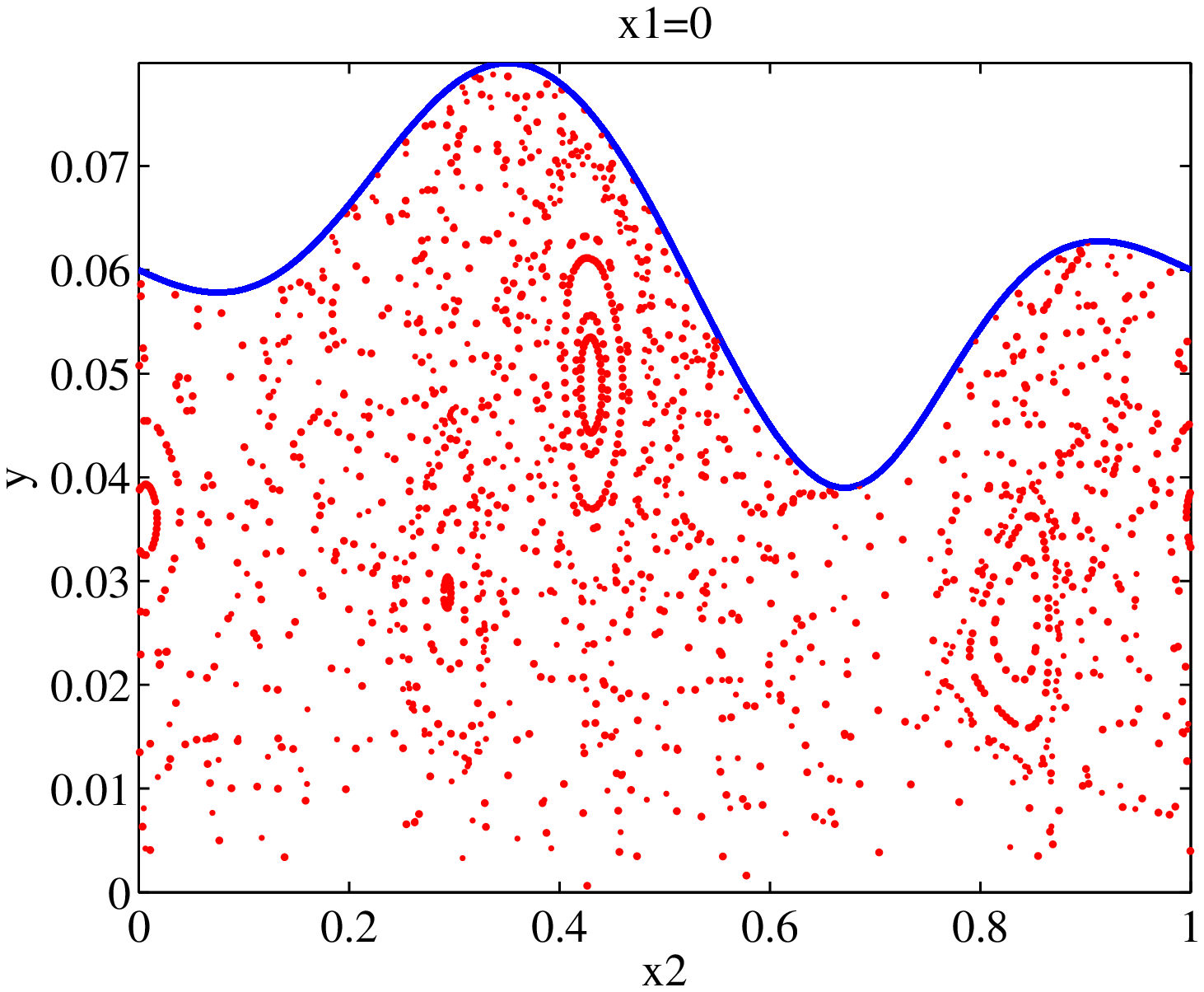}
  \label{fig:psec_2D_chaos}
}\hspace{0em}%
\subfigure[]{
  \psfrag{x2}{$\xc^2$}
  \includegraphics[width=.48\textwidth]{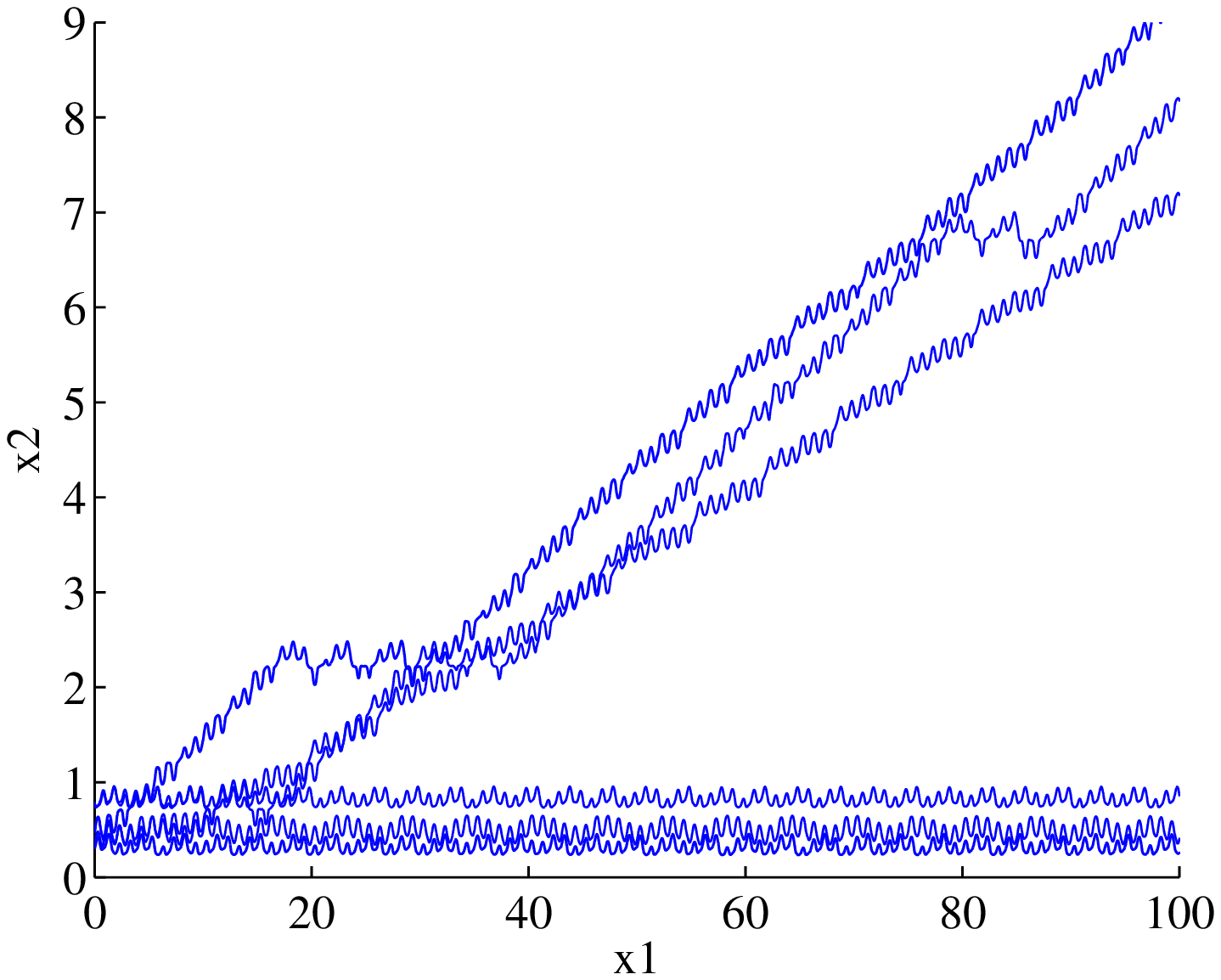}
  \label{fig:traj_2D_chaos}
}
\caption{For same parameters as the flow used for
  Fig.~\ref{fig:nochaos} but with a shallower inclination
  angle~$\theta=0.1$: (a) Poincar\'e section at~$\xc^1=0$, showing
  considerable chaos; (b) Six typical trajectories, showing three trapped
  trajectories and three trajectories that undergo trapping events and flights
  typical of chaotic dynamics.}
\label{fig:chaos}
\end{figure}

Now we apply our third modification from the list above and decrease the
inclination angle to~$\theta=0.1$.  This gives the thickness field depicted in
Fig.~\ref{fig:chaos_sol}.  Figure~\ref{fig:psec_2D_chaos} shows a Poincar\'e
section for the resulting flow.
The phase space clearly exhibits considerable chaotic behavior, as evidenced
by the Poincar\'e section not being foliated by one-dimensional surfaces
[compare to Fig.~\ref{fig:psec_2D_nochaos}].  There are still a few regions of
regular behavior, indicated by the four visible islands, which correspond to
islands in Fig.~\ref{fig:psec_2D_nochaos}.  In Figure~\ref{fig:traj_2D_chaos}
six trajectories are plotted: the bottom three are trapped in the islands.
The top three undergo transitions between long untrapped motions, where the
trajectories go diagonally across the substrate, and short trapped motions.
This trapped-untrapped behavior is typical of chaotic dynamics in the
neighborhood of islands.~\cite{Solomon1993,Solomon1994}

\section{Discussion}
\label{sec:discussion}

We have derived an equation describing the gravity-driven motion of a thin
viscous flow on a curved substrate.  This is the same equation as in
RRS~\cite{Roy2002} expressed in nonorthogonal coordinates, and in addition
including second-order correction terms to satisfy exactly the kinematic
boundary condition at the free surface.  This is essential for particle
advection, since otherwise the small errors at the surface are enough to cause
the particles to escape the domain.

We then investigated the character of fluid trajectories for four cases of
increasing complexity.  We observe a hierarchy of complex motions:
\begin{itemize}
\item For a substrate with translational symmetry at a large inclination
  angle, the flow is very thin and particle moves in uniform layers, as in
  Fig.~\ref{fig:psec_1D_norolls}.  The lateral motion of the particles is
  essentially independent of their depth [Fig.~\ref{fig:traj_1D_norolls}].
\item For the same substrate with a smaller inclination angle, the flow
  becomes thicker over furrows, and zones of recirculation form.  These are
  seen as islands in Fig.~\ref{fig:psec_1D_rolls}.  Two new effects are
  visible in the particles trajectories [Fig.~\ref{fig:traj_1D_rolls}]: some
  are trapped inside the rolls, and others exhibit different rates of lateral
  displacement in~$\xc^2$.  This means that this flow has much better
  transport properties since particles initially near each other will separate
  rapidly (but not exponentially since there is no chaos).
\item If we break the translational symmetry and use a fully two-dimensional
  surface such as~\eqref{eq:chaos_subst}, we observe the formation of more
  islands (four visible in total) of trapped trajectories
  [Fig.~\ref{fig:psec_2D_nochaos}].  The trajectories `bounce off' the various
  islands causing complicated, but regular, oscillations
  [Fig.~\ref{fig:traj_2D_nochaos}].
\item Finally, decreasing the inclination angle of the previous substrate
  preserves the same four islands [Fig.~\ref{fig:psec_2D_chaos}], but now the
  trajectories undergo a succession of random trapped and untrapped events,
  analogous to L\'evy flights observed in experiments in a rotating tank
  [Fig.~\ref{fig:traj_2D_chaos}].~\cite{Solomon1993,Solomon1994} This is the
  most effective transport achieved here, since particles initially near each
  other will separate exponentially, unless they are both inside the same
  island.
\end{itemize}

We used three modifications to the flow to increase the complexity of particle
trajectories: 1. Breaking discrete symmetries of the substrate; 2. Making the
fluid deeper; 3. Making the inclination angle smaller.  Of the three, we
expect the first two to have a similar effect on any other type of thin flow,
say one driven by Marangoni stresses or shear at the free surface.  The last
simply says that decreasing the mean velocity of the flow can make lateral
motions more important, heightening the chance of observing complex dynamics.

The lateral motion (that is, perpendicular to the mean flow direction) of
particles is fairly small.  We have made little effort to optimize this,
preferring to focus our attention on test cases as a first object of study.
Experiments would be desirable here to see if the observed lateral motion is
of the right order of magnitude.

It is tempting to try to compute numerically an effective diffusion
coefficient by simulating large numbers of particle trajectories.  However,
the numerical codes used here are too slow to achieve the high statistics
needed to measure these coefficients with confidence.  Rather, we hope to use
techniques from homogenization theory to calculate the large-scale transport
properties.~\cite{Majda1999}

The generalized coordinate setting introduced in Section~\ref{sec:coords} is
applicable to any set of dynamical equations, and could be used to model flows
over complicated surbstrates in a fairly straightforward manner.  It also
allows for a time-dependent substrate, such as when it grows by
solidification,~\cite{Myers2002} though we have not treated such effects here.

Finally, the substrates used here were periodic in both directions, but it
would be useful to quantify transport for random landscapes, consisting for
instance of imperfections on the surface of a material.

\section*{Acknowledgments}

We thank Richard Craster for his valuable advice.  J-LT was funded in part by
the UK Engineering and Physical Sciences Research Council grant GR/S72931/01.


\end{document}